\documentclass[aps,superscriptaddress,showpacs,floatfix,twocolumn,prd]{revtex4}
\usepackage{epsfig,graphics,color}
\usepackage{graphicx}% Include figure files
\usepackage{dcolumn}% Align table columns on decimal point
\usepackage{bm}% bold math
\usepackage{amsmath}
\usepackage{eucal,eufrak}

\newcommand \tie {{\it i.e.}}

\newcommand \eg {{\it e.g.}}  
\newcommand \f {\not\!}

\newcommand \p {{\prime}}

\newcommand \ra  {\rightarrow}

\newcommand \fp {{\bf p}}

\newcommand \fy {{\bf y}}

\newcommand \vl {\vec{l}}

\newcommand \vp {\vec{p}}

\newcommand \A {\alpha}
\newcommand \B {\beta}
\newcommand \g {\gamma}

\newcommand \D {\Delta}
\newcommand \kd  {\delta}

\newcommand \e {\epsilon}

\newcommand \h {\theta}

\newcommand \sg {\sigma}

\newcommand \md {\mathfrak{D}}

\newcommand \Op {{\mathcal O}}

\newcommand \x {\cdot}
\newcommand \hf {\frac{1}{2}}

\newcommand \lc {\langle}
\newcommand \rc {\rangle}
\newcommand \prt {\partial}
\newcommand \nt {\noindent}

\newcommand \gmn {g^{\mu \nu}}

\newcommand \bvec{\left( \begin{array}{c} }
\newcommand \evec{\end{array} \right)}
\newcommand \tr {\mbox{{\bf Tr}}}
\newcommand \bea{\begin{eqnarray} }
\newcommand \eea{\end{eqnarray} }
\newcommand \nn {\nonumber}
\newcommand \be {\begin{equation}}
\newcommand \ee {\end{equation}}

\newcommand \mbx {\mbox{}}

\newcommand \psibar {\bar{\psi}}
\newcommand \ata {& \times &}
\newcommand {\apa} {& + &}
\newcommand {\aqa} {&=&}
\newcommand \slm {\sum\limits}
\begin{document}

\title{Photon bremsstrahlung and diffusive broadening of a hard jet}

\author{A.~Majumder}
\affiliation{Department of Physics, Duke University, Durham, NC 27708, USA.}

\author{R.~J.~Fries}
\affiliation{Cyclotron Institute and Department of Physics, Texas A\&M University, College Station, TX 77843, USA.}
\affiliation{RIKEN/BNL Research Center, Brookhaven National Laboratory, Upton NY 11973, USA.}

\author{B.~M\"uller}
\affiliation{Department of Physics, Duke University, Durham, NC 27708, USA.}

\date{\today}

\begin{abstract} 
The photon bremsstrahlung rate from a quark jet produced in
deep-inelastic scattering (DIS) off a large nucleus is studied in the collinear limit.
The leading medium-enhanced higher twist corrections which
describe the multiple scattering of the jet in the nucleus
are re-summed to all orders of twist. The propagation of the jet in the absence of further
radiative energy loss is shown to be governed by a transverse momentum
diffusion equation. We compute the final photon spectrum in
the limit of soft photons, taking into account the leading and
next-to-leading terms in the photon momentum fraction $y$.
In this limit, the photon spectrum in a physical gauge is shown to arise from
two interfering sources: one where the initial hard scattering
produces an off-shell quark which immediately radiates the photon and then undergoes 
subsequent soft re-scattering; alternatively the quark is produced on-shell
and propagates through the medium until it is driven off-shell
by re-scattering and radiates the photon. Our result has a simple
formal structure as a product of the photon splitting function, 
the quark transverse momentum distribution coming from a diffusion 
equation and a dimensionless  factor which encodes the effect of the 
interferences encountered by the propagating quark over the length 
of the medium. The destructive nature of such interferences in the small $y$ limit are 
responsible for the origin of the Landau-Pomeranchuck-Migdal (LPM) effect.
Along the way we also discuss possible implications for quark
jets in hot nuclear matter.
\end{abstract}


\maketitle

%%%%%%%%%%%%%%%%%%%%%%%%%%%%%%%%%%%%%%%%%%%%%%%%%%%%%%%%%%
%%%%%%%%%%%%%%%%%%%%%%%%%%%%%%%%%%%%%%%%%%%%%%%%%%%%%%%%%%
%%%%%%%%%%%%%%%%%%%%%%%%%%%%%%%%%%%%%%%%%%%%%%%%%%%%%%%%%%
%%%%%%%%%%%%%%%%%%%%%%%%%%%%%%%%%%%%%%%%%%%%%%%%%%%%%%%%%%
%%%%%%%%%%%%%%%%%%%%%%%%%%%%%%%%%%%%%%%%%%%%%%%%%%%%%%%%%%

 \section{introduction}

%%%%%%%%%%%%%%%%%%%%%%%%%%%%%%%%%%%%%%%%%%%%%%%%%%%%%%%%%%
%%%%%%%%%%%%%%%%%%%%%%%%%%%%%%%%%%%%%%%%%%%%%%%%%%%%%%%%%%
%%%%%%%%%%%%%%%%%%%%%%%%%%%%%%%%%%%%%%%%%%%%%%%%%%%%%%%%%%
%%%%%%%%%%%%%%%%%%%%%%%%%%%%%%%%%%%%%%%%%%%%%%%%%%%%%%%%%%

The modification of high transverse momentum (high $p_T$) 
jets~\cite{quenching,Baier:1996kr,Zakharov:1996fv} as they pass through dense 
matter has achieved  center stage in the 
experimental program at the Relativistic Heavy-Ion Collider (RHIC) at 
Brookhaven National Laboratory (BNL)~\cite{white_papers} and 
is poised for a more comprehensive exploration at the upcoming Large 
Hadron Collider (LHC). The central goal is an quantitative exploration 
of certain properties of the dense matter produced, based on the 
modification encountered by a jet while passing 
through the excited medium.  The benchmark for such observables is provided by 
the modification encountered by a hard jet in cold nuclear matter~\cite{Airapetian:2000ks}.  
The basic partonic processes which lead to the modification of the jet in cold or excited media 
are identical and lead to a formulation in terms of the same correlation 
functions. Experimental observations of the modification in cold and excited matter provide 
numerical values for these correlation functions and as such allow comparisons between 
these two environments. 

The most noticeable modification observed is the loss of forward momentum and 
energy of the hard partons believed to result from a combination of 
elastic~\cite{Qin:2007rn,Braaten:1991jj} and radiative~\cite{quenching} mechanisms.
The experimentally observed effect is the depletion of high momentum hadrons 
produced in the fragmentation of these partons after they escape the medium~\cite{highpt,Airapetian:2000ks}. 
For the light partons, the energy loss by induced gluon 
radiation is believed to be the dominant mechanism. Calculations based on this are 
by now considerably developed. There currently exist four different 
schemes of radiative energy loss: the Arnold-Moore-Yaffe (AMY)~\cite{AMY}, 
the Armesto-Salgado-Wiedemann (ASW)~\cite{ASW},  
Gyulassy-Levai-Vitev (GLV)~\cite{GLV} and the Higher-Twist (HT)~\cite{HT} approach, each 
incorporating a slightly different approximation scheme in the calculation of the 
radiative gluon spectrum (There have also been attempts to calculate both the broadening of a jet and its 
energy loss in a dipole approach~\cite{Kopeliovich:1998nw}).  
The different schemes also utilize different 
definitions of the medium modified fragmentation function and in most cases,  
apparently different models of the medium. 

The AMY and ASW approaches, based on the previous work of 
Baier {\it et. al.} (usually referred to as BDMPS)~\cite{Baier:1996kr},  
are cast in the thick medium approximation, in 
the sense that the hard parton as well as the radiated gluon scatters multiple 
times in the medium and an infinite series of such scatterings has to be 
re-summed~\cite{f1,Salgado:2003gb}. Whereas, the GLV and HT approach have traditionally 
been cast in the thin medium approximation, where the hard parton and radiated gluon scatter 
a finite number of times in the medium, with the effect of each subsequent scattering being incorporated 
order by order in opacity or twist~\cite{Guo:2006kz}.

Some time ago, it was pointed out that the higher twist formalism may also be extended 
to the thick medium limit~\cite{Fries:2002mu}.
In a recent effort, the formulation of jet modification due to multiple scattering 
within a re-summed higher twist approach was begun~\cite{Majumder:2007hx}. 
In that paper, contributions from all twists, which are enhanced by the length of the medium, 
were re-summed to calculate the transverse broadening 
experienced by a hard parton as it traversed a dense medium without radiative emission. 
The resulting \emph{all-twist} 
expression assumed the physically transparent form of a two dimensional 
transverse momentum diffusion 
equation. The diffusion tensor was shown to be related, up to overall constant factors, to the well 
known jet transport coefficient $\hat{q}$~\cite{Baier:1996kr}. It is the object of the current 
manuscript to extend this formulation of \emph{all-twist} jet modification. The specific problem 
being addressed here is the spectrum of real photons radiated from a partonic jet multiply scattering 
off a dense medium. While the photon is assumed to escape the medium without interaction, very unlike 
the case of a radiated gluon, the problem does encode many of the characteristic features of 
gluon radiation such as the Landau-Pomeranchuck-Migdal (LPM) effect which suppresses very collinear 
radiation .

Besides its role as an intermediate step to the problem of energy loss by gluon radiation, hard 
photon bremsstrahlung from a high $p_T$ jet is a problem of sufficient phenomenological relevance in 
it own right. Traditionally, thermal photons and dileptons have been considered as ideal probes of the plasma as they 
escape the plasma after their formation without any further interaction, directly conveying information 
regarding conditions at their origin~\cite{Shuryak:1980tp}. 
Jet medium interactions, where a hard jet undergoes Compton scattering off a constituent of the medium leading to direct photon~\cite{Fries:2002kt} 
and dilepton~\cite{Srivastava:2002ic} production have also been proposed as deep probes of the matter produced: these 
measure the partonic density of the medium at the scale of the hard parton.  
Recently, photon hadron correlations~\cite{Wang:1996yh,Turbide:2005fk,Zakharov:2004bi}  where a hard photon is produced in the course of the 
multiple forward scattering of a hard parton have emerged as an electromagnetic window to the energy loss transport 
parameter $\hat{q}$.
There now exist a series of measurements of photon and high $p_T$ hadron correlations at RHIC. 
These are meant to be used in tandem with single inclusive high $p_T$ hadron suppression~\cite{Majumder:2006we} and 
hadron-hadron correlations~\cite{maj04d} to characterize the 
jet transport parameters of the medium. 
The results of the current manuscript are directly relevant to 
all near-side measurements of such correlations and represent an extra source of jet broadening in 
away side correlations between a trigger photon and a leading hadron.

The results of this manuscript 
may also be easily ported into the calculation of hard photon production from a hot partonic medium. 
Such calculations, nowadays, are mostly carried out within the finite temperature formalism of AMY~\cite{AMY} 
by assuming that the 
medium exists at asymptotically high temperature and invoking the Hard Thermal Loop (HTL) approximation~\cite{Braaten:1989mz}
(for a complete discussion see Ref.~\cite{Kapusta:2006pm} and references therein).  Our reformulation of this problem in the 
HT approach achieves several goals. The AMY scheme invokes the assumption that the medium is a thermalized plasma 
of quarks and gluons at asymptotically high temperature $T$. The HT scheme makes no such approximation regarding 
the medium; the only approximation made is that the medium have a short distance color correlation length. This is a property 
of a multitude of media including the HTL plasma. Hence, the results derived in the current article should be applicable to 
both cases where the HTL approximation is applicable and where it is not (for instance, in the confined environment 
of cold nuclear matter).  
This conjecture may be easily tested by calculating the diffusion tensor in an hot partonic medium and 
promoting the hard parton from a single probe to an ensemble with a thermal distribution. 
In such a comparison, contributions that occur due to the initial parton being off-shell 
due to its formation in a violent collision should be neglected. 

Such comparisons and 
the associated phenomenology will be deferred to a future publication. Here, the focus will lie on a derivation of 
the all-twist re-summed expression for photon bremsstrahlung from a hard parton traversing a dense medium. 
While the remaining sections specifically treat photon production from a hard jet produced in the DIS off a large nucleus, it should 
be pointed out that this is merely an instrument of convenience and the results may be readily generalized to photon production off 
a hard jet in a different environment. 
To allow for such extensions in the future, our derivations are carried out in 
a factorized form, almost independent of the initial source of production of the hard quark.

The paper is organized as follows: in Sect.~II the simplest calculation of photon bremsstrahlung at 
leading twist is carried out. This is solely performed to familiarize the reader with the notation, frame and 
particular choice of gauge used in this manuscript. In Sect.~III, the problem of photon radiation
from a hard parton that scatters a finite number of times in the medium is analyzed and general 
expressions are derived. In Sect.~IV, sums over the various photon production points in the medium are 
carried out. The soft photon approximation is invoked and the twist expansion in cold nuclear matter 
is carried out in terms of a transverse momentum gradient expansion. In Sect.~V, the resulting 
expressions for soft photon production from a hard parton that scatters $n$ times in the medium 
are re-summed to obtain the all-twist photon spectrum. 
Concluding discussions appear in Sect.~VI.

%%%%%%%%%%%%%%%%%%%%%%%%%%%%%%%%%%%%%%%%%%%%%%%%%%%%%%%%%%
%%%%%%%%%%%%%%%%%%%%%%%%%%%%%%%%%%%%%%%%%%%%%%%%%%%%%%%%%%
%%%%%%%%%%%%%%%%%%%%%%%%%%%%%%%%%%%%%%%%%%%%%%%%%%%%%%%%%%
%%%%%%%%%%%%%%%%%%%%%%%%%%%%%%%%%%%%%%%%%%%%%%%%%%%%%%%%%%
%%%%%%%%%%%%%%%%%%%%%%%%%%%%%%%%%%%%%%%%%%%%%%%%%%%%%%%%%%

 \section{Leading  twist  and collinear photon radiation}

%%%%%%%%%%%%%%%%%%%%%%%%%%%%%%%%%%%%%%%%%%%%%%%%%%%%%%%%%%
%%%%%%%%%%%%%%%%%%%%%%%%%%%%%%%%%%%%%%%%%%%%%%%%%%%%%%%%%%
%%%%%%%%%%%%%%%%%%%%%%%%%%%%%%%%%%%%%%%%%%%%%%%%%%%%%%%%%%
%%%%%%%%%%%%%%%%%%%%%%%%%%%%%%%%%%%%%%%%%%%%%%%%%%%%%%%%%%

In this section, the calculation of the photon production rate from a hard parton, produced in the 
DIS off a large nucleus, which subsequently undergoes multiple scattering in the confined nuclear environment is 
begun. 
The photon, not having scattered in the medium after its production, 
escapes the medium unscathed, revealing the nature of the matter at the 
location of its  production. The interference between the photon produced at 
different locations in the hard parton's diffusive history lead to a modulation of the 
factorized rate of photon production.  
In this vein, we focus on an all twist evaluation 
of the differential photon production rate from a hard jet. 
Consider the semi-inclusive 
process of DIS off a nucleus in the Breit frame where  one jet with a transverse 
momentum $l_{q_\perp}$ and a  bremsstrahlung photon with transverse 
momentum $l_\perp$ are produced,  

\bea
{\mathcal L} (L_1) + A(p) \longrightarrow  \mathcal{L} (L_2) + J( l_{q_\perp} ) + \gamma(l_\perp) + X .
\label{chemical_eqn}
\eea

\nt
In the above equation, $L_1$ and $L_2$ represent the momentum of the 
incoming and outgoing leptons. The incoming nucleus of atomic mass 
$A$ is endowed with a momentum $Ap$. In the final state,
all high momentum hadrons ($h_1,h_2,...$) with momenta $p_1,p_2,\ldots$ are detected and their
momenta summed to obtain the jet momentum and $X$ denotes that the 
process is semi-inclusive.

\begin{figure}[htbp]
%\begin{center}
%  \epsfxsize 80mm
%\hspace{0cm}
\resizebox{3in}{1.5in}{\includegraphics[0in,0in][8in,4in]{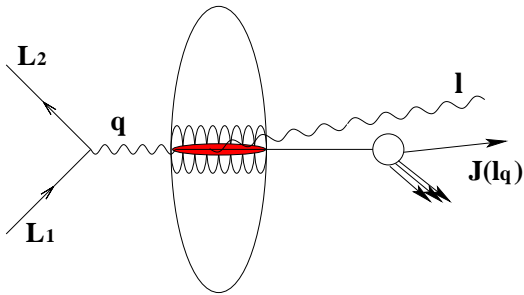}} 
%\vspace{0.25cm}
    \caption{ The Lorentz frame chosen for the process where a nucleon in a large nucleus is
    struck by a hard space-like photon leading to the production of an outgoing parton and a radiated photon.}
    \label{fig1}
%  \end{center}
\end{figure}

Throughout this paper, the light-cone component notation for four vectors ($p \equiv [p^+,p^-, \vec{p}_\perp]$) 
will be used, where, 
\bea
p^+ = \frac{p^0 + p^3}{2}; \,\,\, p^- = p^0 - p^3.
\eea
The kinematics is defined in the Breit frame as sketched in Fig.~\ref{fig1}. 
In such a frame, the incoming virtual photon $\g^*$ and the nucleus have 
momentum four vectors $q,P_A$ given as, 

\[
q = L_2 - L_1 \equiv \left[\frac{-Q^2}{2q^-}, q^-, 0, 0\right], 
\mbox{\hspace{1cm}}
P_A \equiv A[p^+,0,0,0].
\]

\nt
In this frame, the Bjorken variable is defined as 
$x_B = Q^2/2p^+q^-$.  The radiated photon has a transverse momentum of 
$l_\perp$ and carries a fraction $y$ of the forward momentum $q^-$ of the 
quark originating in the hard scattering, \tie, 

\bea
y = \frac{l^-}{q^-} .
\eea

The double differential cross section of the semi inclusive process 
with a jet with transverse momentum $l_{q_\perp}$  and a final 
state photon with transverse momentum $l_\perp$ may be expressed as 

\bea
\frac{E_{L_2} d \sigma } {d^3 L_2 d^2 l_{q_\perp}  d^2 l_\perp dy  } &=&
\frac{\A_{em}^2}{2\pi s  Q^4}  L_{\mu \nu}  
\frac{d W^{\mu \nu}}{d^2 l_{q_\perp} d^2 l_\perp dy}, \label{LO_cross}
\eea

\nt
where $s = (p+L_1)^2$ is the total invariant mass of the lepton nucleon 
system. The reader may have already surmised the form of the leptonic tensor 
as,

\bea 
L_{\mu \nu} = \frac{1}{2} \tr [ \f L_1 \g_{\mu} \f L_2 \g_{\nu}].
\eea

\nt
In the notation used in this paper, $| A; p \rc$ represents the spin averaged initial state of 
an incoming nucleus with $A$ nucleons with a momentum $p$ per nucleon. The 
general final hadronic or partonic state is defined as 
$| X \rc $.
As a result, the semi-inclusive hadronic tensor in the nuclear state $|A\rc$ may be defined as

\bea 
{W^A}^{\mu \nu}\!\!\!\!&=& \!\!\!\! \sum_X  (2\pi)^4 
\kd^4 (q\!+\!P_A\!-\!p_X ) \nn \\
\ata \lc A; p |  J^{\mu}(0) | X  \rc \lc X  | J^{\nu}(0) | A;p \rc  \nn \\
&=& 2 \mbox{Im} \left[  \int d^4 y_0 e^{i q \cdot y_0 } \lc A;p | J^{\mu} (y_0) J^{\nu}(0) | A;p \rc \right].
\eea

\nt
In the equation above,  the sum ($\sum_X$)  runs over all possible hadronic states and $J^{\mu} =  Q_q \bar{\psi}_q \g^\mu \psi_q$ is the 
hadronic electromagnetic current, where, $Q_q$ is the 
charge of a quark of flavor $q$ in units of the positron charge $e$. 
It is understood that the quark operators are written in the  interaction picture, and two 
factors of the electromagnetic coupling constant $\A_{em}$ have already been extracted and 
included in Eq.~\eqref{LO_cross}.
The leptonic tensor will not be discussed further. The focus in the remaining 
shall lie exclusively on the hadronic tensor. This tensor will be expanded 
order by order in a partonic basis with one photon in the final state 
and leading twist and maximally length enhanced higher 
twist contributions will be isolated.

\begin{figure}[htbp]
%\begin{center}
%  \epsfxsize 80mm
%\hspace{0cm}
  \resizebox{3.2in}{1.6in}{\includegraphics[0in,0in][8in,4in]{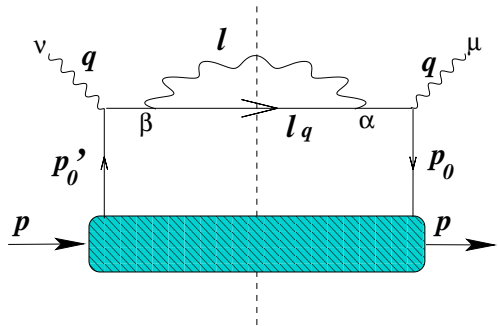}} 
%\vspace{0.25cm}
    \caption{ The Lowest order and leading twist contribution to $W^{\mu \nu}$.}
    \label{fig2}
%  \end{center}
\end{figure}

The leading twist contribution is obtained by expanding the products of 
currents at leading order in $\A_s$ and next-leading-order in $\A_{em}$ to account for the radiated photon. 
This contribution may be expressed diagrammatically 
as in Fig.~\ref{fig2}. This represents the process where a hard quark, 
produced from one nucleon in a deep-inelastic scattering on a nucleus, 
proceeds to radiate a hard photon and then exits the nucleus without further interaction. 
Other diagrams are suppressed in our
choice of light cone gauge. In the following, we analyze this contribution in 
some detail. Indeed, there is no new information presented in the current 
section and the discussion of such contributions is now well established~\cite{gri72}. 
The approximations made in this section will form the 
basis for the analysis of photon production at all twist. The semi-inclusive
hadronic tensor may be expressed as, 

\bea
{W_0^A}^{\mu \nu} 
% &\equiv & 2 Im \left[ \int d^4y e^{iq \x y} \lnuc  J^{\mu}(y) J^\nu (0)  \rnuc \right] \nn \\
%
&=&  A C_p^A W_0^{\mu \nu} \label{w_mu_nu_twist=2}\\
&=& A C_p^A \int d^4y_0 \lc p |  \psibar(y_0) \g^\mu  \widehat{\Op^{00}} \g^\nu \psi(0) | p \rc  \nn \\
&=& C_p^A \int d^4 y_0\tr [\frac{\g^-}{2} \g^\mu \frac{\g^+}{2} \g^\nu ] F(y_0) \Op^{00}(y_0) .\nn 
\eea

\nt
In the above equation, $C_p^A$ expresses the probability to find a nucleon state with momentum $p$ inside a nucleus with $A$ nucleons.

In the collinear 
limit, the incoming parton is assumed to be endowed with very high forward momentum $({p_0}^+ = x_0 p^+, p_0^- \ra 0)$
with negligible transverse momentum ${p_0}_\perp \ll p_0^+$.  Within the 
kinematics chosen, the incoming virtual photon also has no transverse 
momentum. As a result, the produced final state parton also has a vanishingly small transverse momentum 
(\tie, with a distribution $\kd^2(\vec{p}_\perp)$). 
In this limit, the leading spin projection of  the pieces which represent the initial state 
and final state may be taken. The factors, 

\bea
\g^+ = \frac{\g^0 + \g^3}{2} \,\,\, ;\,\,\,\,\,    \g^- = \g^0 - \g^3, 
\eea

\nt
are used to obtain  the spin projections along the leading momenta of the outgoing state and the incoming 
state. The coefficients of these projections are the two functions, 

\bea
F(y_0) = A \lc p | \psibar(y_0) \frac{\g^+}{2} \psi(0) | p \rc \label{F(y_0)}
\eea

\nt and (in a notation where the superscript on the operator $\Op^{00}$ implies that the 
quark undergoes no scattering in the initial or final state)

\bea 
\Op^{00} &=& \tr \left[ \frac{\g^-}{2} \widehat{\Op^{00}} \right]  \label{O_00_1}\\
 &= & \int \frac{d^4 l}{(2\pi)^4}   d^4 z d^4 z^\p 
\frac{d^4 l_q}{(2\pi)^4}\frac{d^4 p_0}{(2\pi)^4} \frac{d^4 p^\p_0}{(2\pi)^4}\nn \\ 
\ata   \tr \left[  \frac{\g^-}{2} \frac{-i (\f p_0 + \f q )}{ (p_0 + q )^2 - i \e }  
i\g^\A  \f l_q   2\pi \kd (l_q^2)  \right.  \nn \\
\ata \left. G_{\A \B} (l)   2\pi \kd (l^2)  (-i\g^{\B})
\frac{ i (\f p_0^\p + \f q )}{ (p_0^\p + q )^2 +  i \e }  \right] \nn \\
\ata e^{i q \x y_0 }e^{ -i (p_0 + q) \x ( y_0 - z) } 
e^{-i l \x (z - z^\p)} e^{-i l_q \x (z - z^\p)} \nn \\
\ata e^{ -i (p_0^\p + q) \x z^\p  } e_q^2  . \nn
\eea

\nt
Integrating over $z$ and $z^\p$ yields the two four-dimensional $\kd$-functions: 
$\kd^4 ( p_0 + q - l - l_q )$ and $\kd^4 (p_0 - p_0^\p )$.

The erudite reader will have noticed that we have ignored various projections such as those which 
arise from the ($\perp$)-components of the $\gamma$ matrices, \eg,

\bea
C_P^A \tr \left[   \frac{\g^{\perp_i}}{2} \g^\mu \frac{\g^{\perp_j}}{2} \g^{\nu} \right] F_{\perp_i} (y_0) \Op^{00}_{\perp_j}.
\eea

\nt
This approximation may be justified in the high energy, collinear limit $l_\perp^2/y << Q^2$ where such contributions are 
suppressed compared to those of Eq.~\eqref{w_mu_nu_twist=2}. In so doing, the focus of the remainder of 
this article has been limited to projections where the incoming virtual photon is transverse. As the primary impetus of the 
current manuscript is a description of the effect of final state scattering on the out-going quark leading to final 
state photon bremsstrahlung, we defer the discussion of the slightly different process, where the initial virtual photon is 
longitudinal, to a separate effort.

The on-shell $\delta$-function over $l$ is used to set $l^+ = l_\perp^2/2l^-$. 
The other on-shell $\kd$-function, instills the condition, 
\bea
\kd(l_q^2) &=& \kd[ ( p_0 + q - l )^2 ]  \nn \\ 
&\simeq& \kd [ -Q^2 + 2p_0^+ (q^- - l^-) - 2q^+ l^- - 2q^- l^+  ]     \nn \\
\aqa \frac{1}{2 p^+ q^- } \kd \left[ x_0 (1 - y) - x_B ( 1 - y  )  - \frac{l_\perp^2}{2 p^+ q^- y } \right] \nn \\
\aqa \frac{\kd [x_0 - x_B - x_L ]}{2 p^+ q^- (1- y)} , \label{delta_l_q}
\eea
 
\nt
where, the collinear condition that $p_{0_\perp} \ra 0$ has been used  to simplify the final 
equation. The new \emph{momentum fraction} $x_L$ has been introduced: 

\bea
x_L = \frac{l_\perp^2}{ 2 p^+ q^- y (1-y) } = \frac{1}{p^+ \tau_f} , 
\eea
where 
$y$ has already been defined as the momentum fraction of the radiated photon ($ l^-/q^-$)
and $\tau_f$ is the formation time of the radiated photon.

The factor $G_{\A \B}(l)$ in Eq.~\eqref{O_00_1} represents the radiated 
photon's spin sum.  In this effort, the light cone gauge ($A^-=0$) will be assumed, 
\tie,
 
\bea
G_{\A \B} (l) &=& -\gmn + \frac{l^\mu n^\nu + l^\nu n^\mu}{ n\x l} , \label{light_cone_polarization}
\eea

\nt
where we have introduced the light cone vector $n \equiv [1,0,0,0]$ which yields 
$l \x n = l^-$. Note that with this choice of gauge, the largest component of the 
vector potentials from the initial states may still be regarded as the ($+$)-components. 

Substituting the above simplifications in  Eq.~\eqref{O_00_1}, 
leads to the simplified form for the final state projection: 

\bea
\Op^{00} &=& \int \frac{d^4 p_0}{(2 \pi)^4} 
\frac{d l^- d^2 l_\perp}{(2 \pi)^3 2l^-} e_q^2 e^{ -i p_0 \x y_0} \nn \\
\ata \tr \left[ \frac{\g^-}{2} 
\frac{\g^+ q^-  }{2p^+q^-( x_0 - x_B - {x_D}_0  - i \e)} \right.  \nn \\
\ata  
\left\{  \g_\perp^\A \g^- ( [x_0 - x_B] p^+  - l^+   )  
\g_\perp^\B (-{g_\perp}_{\A \B})  \right. \nn \\
&-&  \frac{ \g_\perp \x l }{l^-} \g_\perp \x  l    \g^- 
- \g^- \g_\perp \x  l  \frac{ \g_\perp \x  l }{l^-} \nn \\
\apa \left.   \g^- \g^+ (q^- - l^-) \g^- \frac{2l^+}{l^-}  \right\}   \nn \\
\ata \left. \frac{\g^+ q^-  }{2p^+q^-( x_0 - x_B - {x_D}_0  + i \e)}  \right]  \nn \\
\ata 2\pi \frac{\kd [x_0 - x_B - x_L ]}{2 p^+ q^- (1- y)}. \label{O_00_2}
\eea

\nt
The $\delta$-function may be used to carry out the $p_0^+ = x_0 p^+$ integral; 
the absence of  $p_0^- $ and $p_{0_\perp}$ from the integrands 
allows for these integrals to be carried out and constrain the locations 
$y_0^+, y_{0_\perp}$ to the origin. 
Further simplifications may be carried out by noting that $\g^\pm$ anticommutes with $\g_\perp$, while  
$\{ \g^+, \g^- \} = 2 \mbox{\bf 1}$ (\tie, twice the unit matrix in spinor space) and $\{ \g^\pm , \g^\pm \} =0 $. 
Replacing $l^- = q^- y$, in Eq.~\eqref{O_00_2}, one obtains,

\bea 
\Op^{00} &=& \kd(y_0^+) \kd^2(y_{0_\perp}) \int \frac{d y d^2 l_\perp}{(2 \pi )^3 2 y } 
e^{-i (x_B + x_L)p^+ y_0^-} p^+ \nn  \\ 
\ata \frac{e^2 Q_q^2  4 (q^-)^2 }{(2p^+q^-)^2 4 p^+ q^- (1-y) x_L^2 }  \nn \\ 
\ata \left[  2 ( x_L p^+  - l^+ )  
 + \frac{2 l_\perp^2}{l^-} + 2 (1-y) \frac{2 l^+}{y} \right]  \nn \\
\aqa \!\!\! \kd(y_0^+) \kd^2(y_{0_\perp})  \frac{Q_q^2  \A_{em}}{2\pi}  
\!\!\int \!\! \frac{dy dl_\perp^2}{l_\perp^2}  \frac{2 - 2y + y^2}{y}  . \label{O_00_3}
\eea

Reintroduction of the final state projection $\Op^{00}$  
in Eq.~\eqref{w_mu_nu_twist=2}, produces
the well known and physically clear formula 
for the differential semi-inclusive hadronic tensor with single photon 
emission in the final state,

\bea
\frac{{ dW_0^A}{\mu \nu} }{dy d l^2_\perp} &=&C_p^A 2 \pi  \sum_q Q_q^4 f^A_q (x_B+x_L) 
%
% [ g^{\mu -} g^{\nu +} + g^{\mu +} g^{\nu -} - g^{\mu \nu} ]  
(-g_{\perp}^{\mu \nu}) \nn \\ 
\ata  \frac{\A_{em}}{2\pi}   \frac{1}{l_\perp^2}  P_{q\ra q \g }(y). \label{W_mu_nu_2}
\eea

\nt
In the above equation, $f^A_q(x_B)$ represent the parton distribution function of 
a quark with flavor $q$ and electric charge $Q_q$ in units of the electron charge 
$e$, in a nucleus with momentum $Ap$ \tie,

\bea
f^A_q(x_B+x_L)  &=& A \int \frac{d y_0^-}{2\pi} e^{-i(x_B + x_L)p^+ y_0^-} \nn \\
\ata \hf \lc p| \psibar(y_0^-) \g^+ \psi(0) | p \rc .
\eea

\nt
In Eq.~\eqref{W_mu_nu_2}, the factor $P_{q\ra q \g} (y) = (2 - 2y + y^2)/y$ 
is the quark-to-photon splitting function; 
it represents the probability that a quark will radiate a photon which will carry away a fraction 
$y$ of its forward momentum. The projection $g_\perp^{\mu \nu} =g^{\mu \nu}  -  g^{\mu -} g^{\nu +} - g^{\mu +} g^{\nu -}$.

As the parton produced immediately after the hard scattering has a vanishingly small transverse momentum, 
the transverse momentum of final produced quark is simply the 
negative of the photon's transverse momentum, 
\tie, $\vec{l}_{q_\perp} = - \vec{l}_\perp$.  As a result, the differential hadronic tensor for the 
transverse momentum distribution of the final quark is given as

\bea
\frac{d  W_0^{\mu \nu}}{dy dl_\perp^2 d^2{l_q}_\perp } = \frac{ d W_0^{\mu \nu}}{dy dl_\perp^2}
 \kd^2 (\vec{l}_\perp + \vec{l}_{q_\perp} ). \label{d_W_0}
\eea

%%%%%%%%%%%%%%%%%%%%%%%%%%%%%%%%%%%%%%%%%%%%%%%%%%%%%%%%%%
%%%%%%%%%%%%%%%%%%%%%%%%%%%%%%%%%%%%%%%%%%%%%%%%%%%%%%%%%%
%%%%%%%%%%%%%%%%%%%%%%%%%%%%%%%%%%%%%%%%%%%%%%%%%%%%%%%%%%
%%%%%%%%%%%%%%%%%%%%%%%%%%%%%%%%%%%%%%%%%%%%%%%%%%%%%%%%%%
%%%%%%%%%%%%%%%%%%%%%%%%%%%%%%%%%%%%%%%%%%%%%%%%%%%%%%%%%%

 \section{Photon radiation from multiple scattering}

%%%%%%%%%%%%%%%%%%%%%%%%%%%%%%%%%%%%%%%%%%%%%%%%%%%%%%%%%%
%%%%%%%%%%%%%%%%%%%%%%%%%%%%%%%%%%%%%%%%%%%%%%%%%%%%%%%%%%
%%%%%%%%%%%%%%%%%%%%%%%%%%%%%%%%%%%%%%%%%%%%%%%%%%%%%%%%%%
%%%%%%%%%%%%%%%%%%%%%%%%%%%%%%%%%%%%%%%%%%%%%%%%%%%%%%%%%%

In this section, the calculation of the single photon production rate from diagrams which include the 
multiple scattering of the quark on the gluon field within the various nucleons inside the nucleus 
will be carried out. The focus will be on a generic diagram of the type that will be encountered  
in the evaluation of the photon differential rate at all twist. 
In another language, this implies the inclusion of higher twist corrections to 
Eq.~\eqref{W_mu_nu_2}.  
Higher twist contributions are obtained from diagrams  which  include expectation values of 
products containing more partonic operators in the medium~\cite{Qiu:1990xx} 
\eg, the gluon field strength operator product $F^{+ \nu}(y) F^+_{\nu} (0) $. 
While contributions from the expectation of such operators are suppressed by powers of the hard scale 
$Q^2$, a sub-class of these are enhanced in extended media (such as an atomic nucleus) due to the 
longitudinal extent that must be travelled by the struck quark. As in Ref.~\cite{Majumder:2007hx}, only 
such length enhanced higher twist contributions will be isolated and eventually re-summed. 
Issues relating to the generalized factorization~\cite{Qiu:1990xy,col89} of such contributions will not be 
dealt with in this effort; the focus will be to derive the effect of multiple scattering of the hard parton in the nucleus on 
its ability to radiate a collinear photon. 
To obtain higher twist contributions, higher orders need to be included. A diagram with 
$2n$  gluon insertions may contribute to twist $m \leq 2n$. 
In all calculations, the high energy and hence small $g$ limit will be assumed, as a result 
all diagrams containing the four-gluon vertex will be ignored. 
Within this choice of kinematics and gauge there exist a class of diagrams which contain 
three-gluon vertices and receive similar length enhancements. These diagrams involve radiated 
gluons in the final state and contribute to the energy loss of the parent parton. Such 
length enhanced contributions will be considered separately in an upcoming publication.

The aim of this effort is to isolate the higher twist contributions at a given order in coupling 
and twist that carry the largest 
multiple of length $L \sim A^{1/3}$~\cite{lqs}.  In the language of power counting, one looks at the combination 
$\A_s^m L^n$ (usually $n \leq m$) and focuses on diagrams with the maximum $n$.
The multiples of $L$ are obtained by insisting on conditions that lead to the largest number of 
propagators going close to their on-shell conditions. This is easily achieved by the hard parton 
having the maximum number of space like exchanges with the medium.  
The photon is radiated on-shell and final state scatterings of the photon with the rest of the medium which 
involve additional powers of $\A_{em}$ are ignored.  

 \begin{widetext}
\begin{center}
\begin{figure}[htbp]
%  \epsfxsize 80mm
% \mbx\hspace{2cm}
\resizebox{4in}{3in}{\includegraphics[2.5in,0in][10.5in,5in]{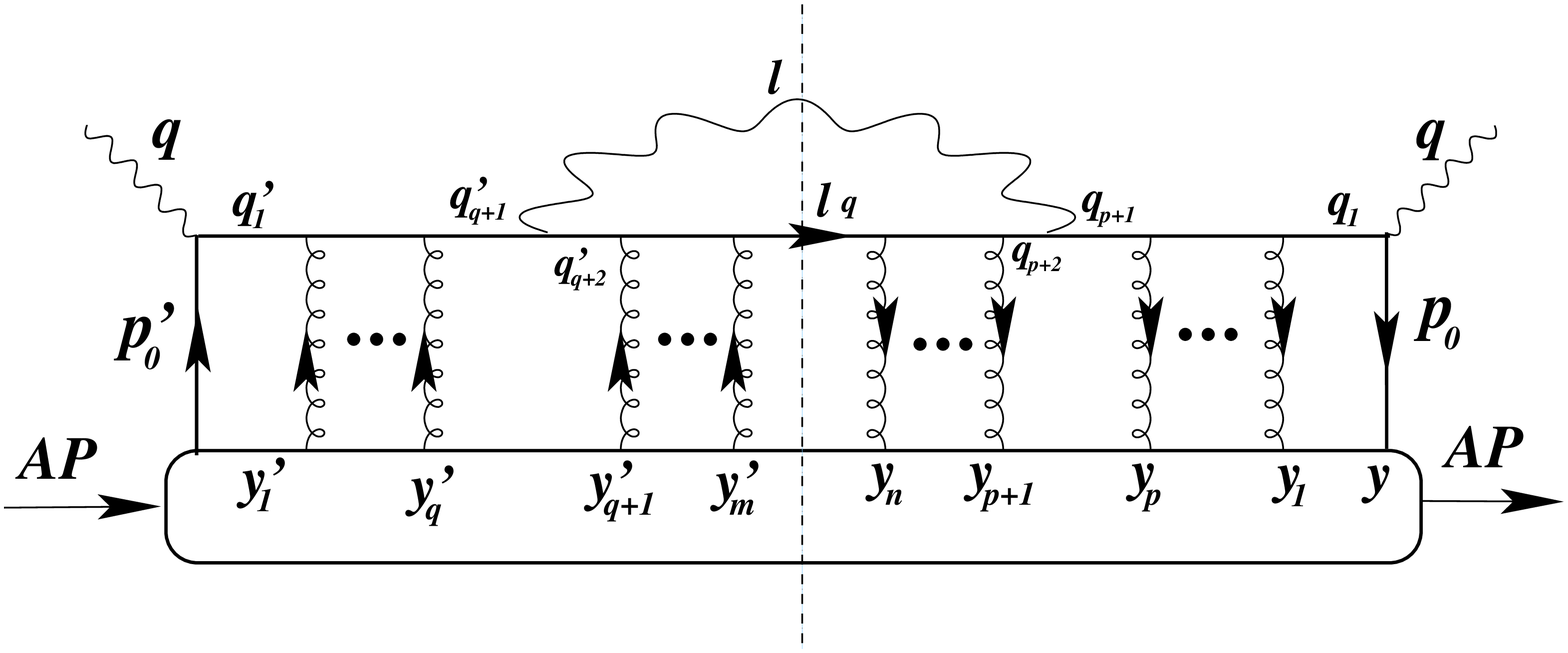}} 
%\vspace{0.25cm}
\caption{
An order $m+n$ contribution to the single photon production rate.  There are $n$ out going 
gluons attached to the quark line on the right hand side of the cut and $m$ incoming gluon lines on the 
left hand side of the cut. The photon attaches between the $p^{\rm th}$ and $(p+1)^{\rm th}$ location on the 
right and between the $q^{\rm th}$ and $(q+1)^{\rm th}$ location on the left. This contributes to twist $k \leq m+n$. }
    \label{fig3}
\end{figure}
\end{center}
\end{widetext}

The generic contributions being considered, thus have the form of Fig.~\ref{fig3}. A hard 
virtual photon strikes a hard quark in the nucleus with momentum $p^\p_0$ ($p_0$ in the complex conjugate) 
at location $y^\p_0 = 0$ ($y_0$ in the complex conjugate) and sends it back through the nuclear medium. 
At this stage, the quark has momentum $q_1^\p$ in the amplitude and $q_1$ in the complex conjugate. 
In the course of its propagation, the hard parton scatters of the gluon field within the medium at locations, 
$y_l^\p$ with $0<l<m$ (at locations $y_i$  with $0<i<n$ in the complex conjugate) wherein the hard parton 
picks up momenta $p_l^\p$ ($p_i$ in the complex conjugate) changing its momentum to $q^\p_{l+1}$ ($q_{i+1}$ in the complex conjugate). 
The photon is radiated between locations $q$ and
$q+1$ in the amplitude and between $p$ and $p+1$ in the complex conjugate. The quark line immediately after the photon radiation 
has momentum $q^\p_{q+2}$ in the amplitude and $q_{p+2}$ in the complex conjugate. 
Following the notation of the 
previous section, the photon has momentum $l$ and the final cut quark propagator has momentum $l_q$.
The ellipses in between 
gluon lines in Fig.~\ref{fig3} are meant to indicate an arbitrary number of insertions. The various momentum equalities outlined 
in this paragraph may be now be expressed succinctly as, 

\bea
q_{i+1} &=& q + \sum_{j=0}^i p_i   \,\,\,\,\,\,  \forall  \,\, [ 0 \leq i \leq p ]   \label{momentum_relations} \\
q^\p_{k+1} &=& q + \sum_{l=0}^k p^\p_k \,\,\,\,\,\,  \forall \,\, [ 0 \leq k \leq q ] \nn \\
q_{i+1} &=& q + \sum_{j=0}^i p_i  - l  \,\,\,\,\,\,  \forall \,\, [ p+1 \leq i \leq n  ]\nn  \\
q^\p_{k+1} &=& q + \sum_{l=0}^k p^\p_k - l  \,\,\,\,\,\,  \forall \,\, [ q+1 \leq k \leq m  ] \nn 
\eea

The general hadronic tensor for such a contribution may be written as  
\bea
&& \left[W^{n,m}_{p,q}\right]_A^{\mu \nu} \label{full_nm_W_mu_nu} \\
% &\equiv & 2 Im \left[ \int d^4y e^{iq \x y} \lnuc  J^{\mu}(y) J^\nu (0)  \rnuc \right] \nn \\
%
&=& AC_{p_1}^A \int d^4y_0 \lc p_1 |  \psibar(y_0) \g^\mu  \left< A \left| \widehat{\Op^{nm}_{p.q}} \right| A \right> \g^\nu \psi(0) | p_1 \rc  \nn .
\eea
In the above equation, we have lumped all operators and propagators except  those of the first struck quark in the amplitude and 
complex conjugate within the multiple operator product $\widehat{\Op^{nm}_{p.q}}$. We have also 
suggestively factorized out one nucleon state denoted as $| p_1 \rc$, which is the nucleon containing the 
struck quark. The remaining nucleons are contained in the state $| A  \rc$~\cite{f2} which is acted upon by the multiple 
operator product $\widehat{\Op^{nm}_{p.q}}$. The coefficient $ C_{p_1}^A$ is meant to indicate the 
correlation between the first nucleon and the rest of the nucleus. We will defer issues related to such a 
factorization of the nuclear state to Sect.~IV where the collinear expansion of the multiple operator 
product will be carried out and the nuclear state decomposed into an ensemble of nucleon states. 
The hadronic tensor, may be further decomposed in terms of the leading spin projections of the 
two operator products, 

\bea
\mbx\!\!\!\!\!\!\left[W^{n,m}_{p,q}\right]_A^{\mu \nu} \!\!\!
&=& \!\!\! C_{p_1}^A\int d^4y_0 \tr [\frac{\g^-}{2} \g^\mu \frac{\g^+}{2} \g^\nu ] F(y_0) \Op^{n,m}_{p.q} ,  \label{hard_fact}
% %
\eea

\nt
where, $F(y_0)$ is defined as in Eq.~\eqref{F(y_0)}  with $|p\rc$ replaced with $|p_1\rc$ and $\Op^{n,m}_{p.q}$ 
is the component of the leading spin projection of the final state operators, 

\bea
\Op^{n,m}_{p,q} &=& \tr \left[ \frac{\g^-}{2} \left< A \left| \widehat{\Op^{n,m}_{p,q}} \right| A \right>  \right] 
\eea

\nt
In the remainder of this section, we will focus exclusively on the component $\Op^{n,m}_{p,q} $.

The expression for $\Op^{n,m}_{p,q} $, may now be written down by applying the Feynman rules to 
the diagram of Fig.~\ref{fig3}. The large rectangular blob is meant to indicate the nuclear state. The lines 
connecting the quark propagators to the blob are not propagators themselves; these are simply quark and gluon 
operator insertions, they introduce a certain momentum, spin and color into the diagram. Note that the quark 
operator insertions that couple with the initial incoming virtual photon have already been extracted from the 
diagram and placed in Eq.~\eqref{full_nm_W_mu_nu}.
Note, there is 
no particular ordering of the gluon lines. 
As these are not propagators there is no meaning associated with 
crossed gluon lines. The entire set of $n + m $ vertex insertions (with the gluon vector potentials 
contracted with the nucleus) 
may then be connected by quark propagators in $(n + m )!$ ways 
(this overall combinatorial factor is removed by
 the $(n + m )!$ which appears in the denominator from 
the perturbation expansion).  All the Feynman propagators are written in position space as Fourier  transforms 
of their momentum space expressions, \eg, for the case of a propagator in the complex conjugate amplitude 
between the $i^{\rm th}$ and $i+1^{\rm th}$ insertion, where $i<p$, we obtain, 

\bea
{ \mathcal T} \left[ \psi(y_i) \psibar(y_{i+1})  \right] &=&
\int  \frac{d^4 q_{i+1}}{(2\pi)^4} \frac{i \f q_{i+1} e^{-i q_{i+1} \x (y_i - y_{i+1}) }  }{q_{i+1}^2 - i \e} 
\nn  \\
\aqa  \int \frac{d^4 p_{i}}{(2\pi)^4} \frac{i ( \f q +\sum_{j=0}^i \f p_{j} ) }{(q + \sum_{j=0}^i p_{j})^2 - i \e} \nn \\ 
\ata e^{-i \left( q+\sum_{j=0}^i p_j \right) \x (y_i - y_{i+1}) }. 
\eea

Using the remaining relations between the various momenta as outlined in Eq.~\eqref{momentum_relations} the 
expression for the leading spin component of final state operator product may be expressed as, 

\begin{widetext}

\bea
\Op_{pq}^{nm} &=& \prod_{i=1}^{n} \prod_{k=1}^{m} d^4 y_i d^4 y^\p_k 
\frac{d^4 p_{i-1}}{(2\pi)^4} \frac{d^4 p^\p_{k-1}}{(2\pi)^4} \frac{d^4 l}{(2\pi)^4}  \frac{d^4 l_q}{(2\pi)^4} 
\label{O_00_4} \\
\ata 
\tr \left[ \frac{1}{2} 
\prod_{i=0}^{p} \left\{  \g^- \frac{ (\sum_{j=0}^{i} \f p_j) + \f q }{ [   ( \sum_{j=0}^{i} p_j  )  +  q  ]^2 - i\e} 
 \right\} \right. 
%\nn \\
%
%\ata 
 \g^\A 
\prod_{i=p}^{n-1} 
\left\{  \frac{   ( \sum_{j=0}^{i} \f p_j )   + \f q - \f \,l }{ [    ( \sum_{j=0}^{i} p_j   ) + q - l   ]^2 - i\e} 
\g^- \right\} \nn \\
\ata 2\pi \kd(l^2) 2\pi \kd(l_q^2) G_{\A \B}(l) \f\,l_q \nn \\
\ata \prod_{k=m-1}^{q} 
\left\{ \g^- \frac{  ( \sum_{l=0}^{k} \f p^\p_l  )  + \f q - \f \,l }{ [     (  \sum_{l=0}^{k} p^\p_l  )   + q -  l ]^2 + i\e} 
 \right\} 
%\nn \\
%
%\ata
\g^\B 
\prod_{k=q}^{1} \left\{  \frac{ (  \sum_{l=0}^{k} \f p^\p_l  )  + \f q }{ [   (  \sum_{l=0}^{k} p^\p_l   )  + q ]^2 + i\e} 
\g^- \right\} 
%\nn \\
%
% \ata
 \left. \frac{ \f p^\p_0 + \f q }{ (  p^\p_0 + q)^2 + i\e} \right]  \nn \\
\ata \prod_{i=0}^{n-1} e^{-ip_i \x y_i} \prod_{ k=0 }^{m-1} e^{ i p^\p_k \x y^\p_k } 
%\nn \\
%
% \ata 
e^{-i y_n \x \left\{ l_q - \left(  \sum\limits_{i=0}^{n-1} p_i \right)  +  q -  l  \right\} } 
e^{ i y^\p_m \x \left\{ l_q - \left(   \sum\limits_{k=0}^{m-1} p^\p_k \right)  + q - l   \right\} }  \nn \\
% \nn \\
%
% \ata 
\ata \left\lc A \left| \prod_{i=1}^n t^{a_i} A^+_{a_i} (y_i) 
\prod_{k=m}^1 t^{b_k} A^+_{b_k} (y^\p_k) \right| A \right\rc.  \nn
\eea

\end{widetext}

The expression for $\Op_{pq}^{nm}$ in Eq.~\eqref{O_00_4}  may be decomposed into a formal convolution of three terms, 

\bea
\Op_{pq}^{mn} &=& \int d^4 l  d^4 l_q \md y \md y^\p \md p \md p^\p \nn \\
\ata T(\fp, \fp^\p, q, l, l_q) \Gamma(\fp,\fp^\p,\fy,\fy^\p) M(\fy,\fy^\p)  \label{general_form}
\eea
\nt
where, $T$ denotes the trace in Eq.~\eqref{O_00_4}, which is solely a function of  the momenta,  
$M(\fy, \fy^\p) = \lc A | \cdots |A \rc$  is the pure position dependent multi-operator 
matrix element and 
$\Gamma (\fp,\fp^\p,\fy,\fy^\p)$ is the phase factor which contains both  
positions and momenta. The bold face quantities $\fp, \fy, \fp^\p, \fy^\p$ represent 
an array of momenta and positions, 

\[
\fp \equiv [ p_0,\ldots,p_{n-1}] ;\,\, \fp^\p \equiv [p^\p_0,\ldots,p^\p_{m-1}  ] ,
\]

\[
\fy \equiv [y_0,\ldots,y_n] ; \, \, \fy^\p \equiv [y^\p_1,\ldots,y^\p_m].
\]  
  
\nt
The range of locations in the complex conjugate amplitude start from $y_0$, 
which is the location of the initial struck quark, whereas, the location of the 
same parton in the amplitude is at the origin. Hence $y_0^\p = 0 $ is not 
explicitly written in the equations presented. 
The integration measures $\md p, p^\p$ and $\md y, y^\p$ denote a 
product of integrals over the different four vectors 
contained in the arrays above.

In an effort to simplify the writing of the matrix element and the momentum dependent part, 
a certain aspect of collinear dynamics has been already been introduced.
As calculations will be carried out in the light cone gauge ($A^- = 0$) in the Breit frame at very high 
energy, the dominant components of the vector potential are the forward or 
$(+)$-components~\cite{lqs}, \tie, $A^\sg \sim g^{- \sg} A^{+}$. 
As  a result, the corresponding 
$\g$ matrices have solely a $(-)$-component. 

To simplify the phase factor, an $n^{\rm th}$ momentum ($p_n$) may be introduced, with  
\begin{equation}
1 = \int d^4 p_n  \kd^4 \left( l_q - \sum_{k=0}^{n} p_k - q + l \right). \label{pn_delta}
\end{equation}
\nt
This leads to a considerable simplification of the phase factor as 

\begin{eqnarray}
\Gamma &=& \exp\left[ - \sum_{i=0}^n i p_i \x y_i + 
\sum_{j=0}^{m - 1} i p^\p_j \x y^\p_j \right. \nn \\ 
&+& \left. i y^\p_{m} \x \left( \sum_{i=0}^{n} p_i - 
\sum_{j=0}^{m-1} p^\p_j \right) \right]. \label{simple_phse}
\end{eqnarray}
\nt
Note the complete absence of the initial hard photon momentum $q$, or the final 
state momenta $l,l_q$ 
from the phase factor. The integrations over the 
lightcone components of the cut line $l_q$ have been performed using two of the four 
$\delta$-functions introduced in Eq.~\eqref{pn_delta}. The remaining two components of the
transverse momentum integration ($d^2 l_{q_\perp}$) are constrained by the$\delta$-function,

\bea
\kd^2(\vec{l_q}_\perp + \vec{l}_\perp - \vec{K}_\perp) 
&=&  \kd^2 \left( \vec{l}_{q_\perp} + \vec{l}_\perp - 
\slm_{i=0}^n \vec{p}^{\,i}_\perp    \right) ,
\eea
where, $\vec{K}_\perp$ is a representative of the sum of the transverse momenta 
brought in by the $n$ gluon insertions.

The approximations stemming from collinear dynamics may 
now be instituted to further simplify the momentum dependent 
coefficient $T$ in Eq.~\eqref{general_form}.  The calculation is carried out in the 
Breit frame at very high energy. As a result, all momentum 
lines that originate in the target are dominated by the 
large ($+$)-components of their momentum, followed by their 
transverse coordinates, \tie,

\bea
p_i^+ \gg {p_i}_\perp \gg p_i^-.
\eea
\nt
In most cases the above condition will allow us to practically 
drop all ($-$)-components of momentum from for the 
hadronic tensor and focus solely on the ($+$) and ($\perp$)-components.
It should be pointed out that the 
($-$)-components are only being dropped from locations where they 
appear in addition to the larger ($+,\perp$)-components. These 
are not dropped from the phase factors in $\Gamma$. 

The simplification of the numerator of $T$ begins by isolating the largest 
components of the momentum. In most cases, this essentially reduces to 
retaining the sole factor $\g^+ q^-$ in the numerator of each of the 
propagators. 
A complication arises from the presence of a radiated photon in the 
final state. The structure of the sum over polarizations controls which 
terms are to be retained. It should be pointed out that unlike the sum over 
polarizations in covariant gauge, 

\bea 
G^{+-} = G^{-+} = 0; \, \, G^{++} = \frac{2l^+}{l^-}; \,\, G^{\perp +} = G^{+ \perp} = \frac{l^\perp}{l^-} ,
\eea

\nt
while, $G_{\perp}^{\A \B} = - g_{\perp}^{\A \B}$ as in the covariant gauge. 
Here and in what follows, the $\perp$ tensor notation is introduced, \tie, $A_{\perp}^{\A \B \cdots}$ 
indicates that the only 
non-zero component of the tensor $A$ involve its transverse components. The 
Dirac trace in the numerator of Eq.~\eqref{O_00_4} denoted by $N(T)$ may be separated 
from its propagation structure involving the denominator of the propagators and the $\delta$-functions denoted as 
$D(T)$. These may then be evaluated separately, as in the following. 

The approximation is begun by ignoring all appearances of the ($-$)-components of the momentum
which originate in the nuclear state \tie, $p_i^-, {p_l^\p}^-$, from all terms in the purely momentum dependent 
component of the integrand $T$,  These momenta now appear solely in the phase factors, which allows 
for the $p^-$ and ${p^\p}^-$ integrations to be done, resulting in the localization of the 
process on the negative light-cone, \tie, 

\bea
\Gamma^- &=&
\prod_{i=0}^n \prod_{l=0}^{m - 1} \int  d{p^\p}_l^- d p_i^-
e^{-\slm_{i=0}^n ip_i^- (y_i^+ - {y^\p}^+_m ) } \nn \\
\ata e^{\slm_{l=0}^{m - 1} i{p^\p_l}^- (y_l^+ - {y^\p}^+_m  ) } \nn \\
&=& \prod_{i=0}^n \kd( y_i^+ - {y^\p}^+_{m} )
\prod_{l=0}^{m - 1} \kd( {y^\p}_l^+ - {y^\p}^+_{m} ). 
\eea

\nt
In the above equation, ${y^\p}_0 = 0$ and as may be noted from the 
definition of the location arrays is not being integrated over. 
As a result, this 
constrains all the negative light cone locations in the 
above equation to the origin.

A set of new quantities may now be introduced by organizing the 
various factors which appear in the denominators of  Eq.~\eqref{O_00_4}. There remain 
the obvious longitudinal momentum fractions, 

\bea
Q^2 = 2x_B p^+ q^-\,; \,p_i^+ = x_i p^+  \,;\, {p^\p}^+_j = x^\p_j p^+ . \label{long_def}  
\eea
As in Ref.~\cite{Majumder:2007hx}, 
we introduce two sets of momentum fractions which are dependent on the transverse momenta imparted 
to the struck quark from the medium, 
\bea
{x_D}_i  = \frac{ \sum_{j=0}^{i} 2 {\vp_\perp^{\,i}} \cdot \vp_\perp^{\,j} + |{\vp_\perp^{\,i}}|^2 }{2p^+ q^-};  \label{i_perp_def} 
\eea
\bea
{x^\p_D}_k = 
\frac{ \sum_{l=0}^{k} 2 \vec{p^\p}_\perp^{k} \cdot \vec{p^\p}_\perp^l + |\vec{p^\p}_\perp^k|^2}{2p^+ q^-}. 
\label{j_perp_def}  
\eea
Two new sets of momentum fractions which depend on both the transverse momentum imparted from the medium as 
well as that carried out by the radiated photon are also introduced:
\bea
{x_L}_i 
= \frac{|\vl_\perp|^2 - y\slm_{j=0}^{i} 2\vl_\perp\cdot \vp_\perp^{\,i} }{ 2 y (1-y) p^+q^- };  \label{l_i_perp_def}
\eea
%\\
\bea
{x^\p_L}_k = \frac{ |\vl_\perp|^2 - y\slm_{l=0}^{k} 2\vl_\perp\cdot \vec{p^\p}_\perp^l }{ 2 y (1-y) p^+q^-} ;  \label{l_j_perp_def} 
\eea

\nt
In the above equation, $i$, the index of the unprimed momenta 
(both for the longitudinal and transverse components) runs from $0$ to $n$, 
whereas, $k$, the index of the primed momenta runs from $0$ to $m - 1$, \tie, 
one less than maximum. The other indices, $j$ and $l$ denote smaller ranges of sums, with $0<j<i$ and $0<l<k$. 
The reader will note the obvious role played by the factors above in the limit that the ($-$)-components 
of all momenta which originate in the nucleus have been ignored and one expands around the dominant 
($+$)-components along with the vectors $q$ and $l$ which have large ($-$)-components. In this limit 
one may factor out the large components of the momentum \tie, $p^+,q^-$ and express the remaining 
components such as combinations and products of $\vl_\perp$, $\vp^{\,i}_\perp$ and $\vec{p^\p}^k_\perp$ as 
fractions of the large product $p^+q^-$ such as ${x_L}_i$, ${x_D}_i$ and ${x_D^\p}_k$.  

Invoking the notion of the collinear approximation, the various definitions in Eqs.~(\ref{long_def}-\ref{l_j_perp_def})
may be used to simplify the integrations that need to be performed in Eq.~\eqref{O_00_4}. the factors 
appearing in $D(T)$ may be written in a simplified form along with the longitudinal 
part of the phase factor \eqref{simple_phse} as, 
\begin{widetext}
\bea 
D(T) &=& \prod_{i=0}^{p}  
\left[ 2p^+ q^-\left( \sum_{j=0}^i  x_j - {x_D}_i - x_B - i\e \right) \right]^{-1} 
%
%\nn \\
%
% \ata
%
 \prod_{i=p}^{n-1}
\left[ 2p^+ q^- (1-y) 
% \right. 
%
% \nn \\ 
%
% \ata 
% \left.
\left( \sum_{j=0}^i  x_j - \frac{{x_D}_j}{1-y} - x_B - {x_L}_i - i\e \right)  \right]^{-1} \nn \\
\ata \frac{ \kd \left( \sum\limits_{j=0}^n  x_j - \frac{ {x_D}_j } {1-y} - x_B - {x_L}_n \right)  } {2p^+q^- (1-y)}     \nn \\
\ata \prod_{k=m-1}^{q} \left[ 2p^+q^- (1-y) 
% \nn \\
%
% \ata \left. 
\left( \sum_{l=0}^k x^\p_l - \frac{  {x^\p_D}_l }{ 1-y }  - x_B - {x^\p_L}_k + i \e \right) \right]^{-1} 
%\nn \\
%
% \ata 
\prod_{k=q}^{0} \left[  2 p^+ q^-  \left( \sum_{l=0}^k   x^\p_l - {x^\p_D}_l - x_B + i\e \right)   \right]^{-1} \nn \\
\ata \exp\left[ - \sum_{i=0}^n i x_i p^+ \x y_i^- + 
\sum_{l=0}^{m - 1} i x^\p_l p^+ \x {y^\p_l}^- 
% \right. \nn \\ 
%
% &+& \left. 
+ i {y^\p_{m}}^- \x p^+ \left( \sum_{i=0}^{n} x_i - 
\sum_{l=0}^{m-1} x^\p_l \right) \right] \label{D(T)}
\eea
% \end{widetext}

\nt
In the above equation, the integration over $x_n$ may be performed with the $\delta$-function which denotes the 
cut quark line to obtain,  
\bea
x_n = x_B + {x_L}_n + \sum_{i=0}^n {x_D}_i - \sum_{i=0}^{n-1} x_i. \label{eql_1}
\eea

\nt
This may be used to rearrange the  longitudinal part of the phase factor as,  

\bea
\!\!\!\!\!\Gamma^+ &=& \exp \left[ -i\left( x_B + {x_L}_n + 
%\\
\frac{\slm_{i=0}^n {x_D}_i }{1-y} \right) p^+ (y_n^-  - {y^\p}_m^-  )\right] 
%\nn \\
%
% \ata 
\prod_{i=0}^{n-1} e^{\left[ -i  x_i p^+ (y_i^-  - y_n^-  )\right] }
% \nn \\
%
% \ata 
\prod_{l=0}^{m-1} e^{\left[ i  {x^\p}_l   p^+ ({y^\p}_l^-  - {y^\p}_{m}^-  )\right]},
\eea
\nt
where, the second line in the equation above involves only momentum fractions and 
locations from the left-hand side of the cut, where as the last line involves momentum 
fractions and locations from the right hand side of the cut line. 
The integrations over the remaining longitudinal momentum fractions, may now be performed 
starting from the propagators adjacent to the cut and proceeding to the propagators adjacent to 
the photon vertices.  

The first such integration, involves the propagator from the 3$^{\rm rd}$ line of Eq.~\eqref{D(T)}. 
Isolating the piece that depends on the fraction $x_{n-1}$ yields the integral, which may 
be performed by closing the contour of $x_{n-1}^-$ with a counterclockwise semi-circle in 
the upper half of the complex plain of,  

\bea
&& \int \frac{dx_{n-1}}{2\pi} 
\frac{e^{-ix_{n-1} \x p^+ (y_{n-1}^- - y_n^- ) }} 
{ x_{n-1} + \slm_{i=0}^{n-2} (x_i  - \frac{{x_D}_i}{1-y})  
- \frac{{x_D}_{n-1}}{1-y}  - x_B - {x_L}_{n-1} - i\e } \nn \\
&=&  i  \h(  y_n^- - y_{n-1}^- )  
e^{  -i \left[  \frac{ {x_D}_{n-1} }{1-y}  + x_B + {x_L}_{n-1} 
- \slm_{i=0}^{n-2} (x_i  - \frac{{x_D}_i}{1-y})   \right]    
p^+ (y_{n-1}^- - y_n^- ) }. 
\eea
\nt
The effect of performing the above integration is the incorporation of 
both the real and imaginary parts of the above propagator into the 
overall expression of the hadronic tensor. It has the physical effect of 
propagating the quark from $y_{n-1}^-$ to $y_n^-$. 
Similarly, the integration over the propagator to the immediate right of the cut line 
may be carried out by closing the contour of $x^\p_{m-1}$ with a clockwise semi-circle in 
the lower complex plain, : 

\bea
&& \int \frac{dx^\p_{m-1}}{2\pi}  
 \frac{e^{ix^\p_{m-1} p^+ ({y^\p}_{m-1}^- - {y^\p}_{m}^- ) }} 
{ x^\p_{m-1} + \slm_{j=0}^{m - 2} ( x^\p_i  
- \frac{{x^\p_D}_i }{1-y})  - \frac{{x^\p_D}_{m-1}}{1-y}  - x_B  - {x^\p_L}_{m-1}+ i\e } \nn \\
&=& -  i  \h(  {y^\p}_{m}^-  - {y^\p}_{m-1}^- ) 
e^{ i \left[ \frac{{x^\p_D}_{m-1}}{1-y}  
+ x_B + {x^\p_L}_{m-1}- \slm_{j=0}^{m-2} ( x^\p_i  - \frac{{x^\p_D}_i}{1-y} )    \right]    
p^+ (  {y^\p}_{m-1}^-    -     {y^\p}_{m}^- ) } . 
\eea

Incorporation of the results of the above two  integrals into the longitudinal 
phase factors leads to the expression, 

\bea
\Gamma^+ &=& \exp\left[   -i \left( \frac{{x_D}_n - \kd {x_L}_n  }{1-y}   \right)p^+  y_n^-    
+  i \left( \frac{ {x^\p_D}_m - \kd {x^\p_L }_m } {1-y}  \right)p^+ {y^\p}^-_m\right]  \nn \\
\ata \exp \left[  -i \left( x_B  + {x_L}_{n-1}  
+ \frac{ \sum_{i =0}^{n-1} {x_D}_i }{1-y} \right) p^+ y_{n-1}^-   
%\right. \nn \\ 
%
% \apa  \left. 
+ i  \left( x_B  + {x_L}^\p_{m-1} 
+ \frac{  \sum_{l=0}^{m - 1} {x^\p_D}_l }{1-y}   \right) p^+ {y^\p}_{m - 1}^-  \right] \nn \\
\ata \prod_{i=0}^{n-2} \exp \left[ -i  x_i p^+ (y_i^-  - y_{n-1}^-  ) \right] 
% \nn \\
%
% \ata
 \prod_{l=0}^{m-2} \exp \left[ i  {x^\p}_l   p^+ ({y^\p}_l^-  - {y^\p}_{m-1}^-  )\right], \label{phase_n-1}
\eea

\end{widetext}

\nt
where, the notion of overall momentum conservation was invoked to define the new variable 
$\vec{p^\p}^m_\perp = \sum_{i=0}^n \vp^{\,i}_\perp - \sum_{k=0}^{m-1} \vec{p^\p}^k_\perp$ and as a result,

\bea
{x^\p_D}_{m} = \slm_{i=0}^{n} {x_D}_i  -  \slm_{j=0}^{m - 1 } {x^\p_D}_j. 
\eea

\nt
In this equation, the quantities labeled ${x_D}_i$ for $0 < i < n-1 $ are defined in 
Eq.~\eqref{i_perp_def}, and ${x^\p_D}_j$ for $0<j< m -1$ are defined in 
Eq.~\eqref{j_perp_def}. The $n^{\rm{th}}$ transverse fraction ${x_D}_n$ is set by 
the $\delta$-function arising from the cut line and is given as in Eq.~\eqref{eql_1}.
The extra momentum fractions  $ \kd {x_L}_n , \kd {x_L^\p}_m $ arise form the fact that 
there is a small difference between ${x_L}_n $ and ${x_L}_{n-1}$ \tie, 

\bea
{x_L}_n - {x_L}_{n-1} = \frac{-  2 \vec{l}_\perp \x \vp^{\,n}_\perp}{2 p^+ q^-  (1-y)} 
= \frac{ - \kd {x_L}_n}{1-y} 
\eea

\nt
and similarly, 

\bea
{x^\p_L}_m - {x^\p_L}_{m-1} =  \frac{ - 2 \vec{l}_\perp \x \vec{p^\p}^m_\perp}{2 p^+ q^- (1-y)} 
= \frac{- \kd {x^\p_L}_m }{1-y}.
\eea

\nt
It should be pointed out that $\kd {x_L}_n$ or the primed momentum fraction 
has no dependence on the momentum fraction of the radiated gluon $y$. Thus 
such contribution will be retained even in the small $y$ approximation. 

The evaluation of the remaining longitudinal momentum fraction integrals is analogous up to the 
outgoing photon vertices.  Therefore, the general 
result for the integrations over the remaining momentum fractions in the phase factor 
may be carried out up to the integral involving $x_{p+1}$ and $x_{q+1}$.  Around 
the photon insertion there are two propagators, either of which may be evaluated to 
obtain the conditions on $x_p$ and $x^\p_q$. We focus on the $x_p$ integral, the case 
for $x^\p_q$ is completely analogous. The integral in question is 

\begin{widetext}

\bea 
\mbx\!\!\!\!\!\!\! && \int \frac{dx_{p}}{2\pi} 
\frac{e^{-ix_{p}  p^+ (y_{p}^- - y_{p+1}^- ) }} 
{ \left[ x_{p} + \slm_{i=0}^{p-1} (x_i  - \frac{{x_D}_i}{1-y})  
- \frac{{x_D}_{p}}{1-y}  - x_B - {x_L}_p - i\e \right]
% \nn \\
% \ata 
\left[  x_{p} + \slm_{i=0}^{p-1} ( x_i  - {x_D}_i ) 
 - {x_D}_p - x_B - i\e \right]    }  \nn \\
\mbx \!\!\!\! \aqa \!\!\! \frac{ i \h ( y_{p+1}^- - y_p^-)}  { {x_L}_p +  \frac{y}{1-y} \sum_{j=0}^p {x_D}_j   } 
% \nn \\ 
%
% \ata 
\left[  e^{ -i \left\{   \frac{ {x_D}_p }{1-y} 
+ x_B + {x_L}_p - \slm_{i=0}^{p-1} \left(  x_i - \frac{{x_D}_i}{1-y} \right)  
\right\} p^+ (y^-_p - y^-_{p+1}) }  
%  \right. \nn \\
%
% & - & \left.  
- e^{  -i \left\{   {x_D}_p  + x_B 
- \slm_{i=0}^{p-1} \left(  x_i - {x_D}_i  \right) \right\} p^+ (y^-_p - y^-_{p+1}) }   \right]. \label{photon_vertex_props}
\eea

\nt
The origin of two separate contributions lies in the fact that the leading length enhancement arises 
when at most one of the propagators is off-shell. This necessarily has to be one of the 
propagators adjacent to the outgoing photon line. The remaining integrations over the 
the propagators between the photon insertion and the hard scattering vertex, are similar to 
the case of transverse broadening~\cite{Majumder:2007hx}.  

The reader will have noticed that the integrals above are  over the momentum fractions 
$x_i$ and $x^\p_l$, whereas, in Eq.~\eqref{O_00_4}, the integrals are over the momenta,
$p_i^+$ and ${p^\p}_l^+$. We have substituted the definitions of Eq.~\eqref{long_def} and 
hence focus on the momentum fractions instead. Thus each integral produces a dimensionful factor 
$p^+$. There being $n+m+1$ integrals over the ($+$)-components of the momentum 
produces an overall factor of $(p^+)^{n+m+1}$. This factor will not be explicitly written out 
in the remaining simplifications of the different parts of $\Op_{p,q}^{n,m}$ that follow. They 
will be reintroduced in the end of the section, when the different pieces are recombined to 
write down the simplified structure of $\Op_{p,q}^{n,m}$.

Invoking the above simplifications, the factor $D(T)$, \tie,  the part of  $\Op_{p,q}^{n,m}$  
which depends solely on the denominators of the propagators of the hard parton and the 
longitudinal phase factors in Eq.~\eqref{D(T)} takes the form,

% \begin{widetext}
\bea 
D(T) \aqa  \frac{\h (y_n^- > y_{n-1}^- > \cdots > y_0^- )}{  (2p^+q^-)^{n+1}  (1-y)^{n-p} }
\prod_{i=0}^{p}  e^{-i {x_D}_i p^+ y_i^-  }
%  \nn \\
%
% \ata 
\prod_{i=p+1}^{n} e^{-i  \left( \frac{{x_D}_i - \kd {x_L}_i  }{1-y}   \right)p^+  y_i^-    } 
\frac{1}{ {x_L}_p +  \frac{y}{1-y} \sum_{j=0}^p {x_D}_j}  \nn \\
\ata \left[  e^{  -i \left( {x_L}_p   
+ \frac{y}{1-y} \slm_{i=0}^{p} {x_D}_i \right) p^+ y_p^-} 
% \right. \nn \\
%
% &-& \left. 
- e^{  -i \left( {x_L}_p  + \frac{y}{1-y} \slm_{i=0}^p  {x_D}_i \right)p^+ y_{p+1}^- }    \right] \nn \\
\ata \frac{ \h ( {y^\p}_m^- > {y^\p}_{m-1}^- > \cdots > {y^\p}_0^- )}{  (2p^+q^-)^{m+1}  (1-y)^{m-q}  } 
 \prod_{k=0}^{q}  e^{ i {x^\p_D}_k p^+  {y^\p}_k^-  } 
% \nn \\
%
% \ata 
\prod_{k=q+1}^{m} e^{ i  \left( \frac{{x^\p_D}_k - \kd {x^\p_L}_k  }{1-y}   \right)p^+  {y^\p}_k^-    } 
% \nn \\
%
% \ata 
\frac{1}{ {x_L^\p}_q +  \frac{y}{1-y} \sum_{l=0}^q {x^\p_D}_l}  \nn \\
\ata \left[  e^{  i \left( {x^\p_L}_q  + \frac{y}{1-y} \slm_{l=0}^{q} {x^\p_D}_l  \right)   p^+ {y^\p}_q^-  } 
% \right. \nn \\
%
% &-& \left. 
-  e^{  i \left( {x^\p_L}_q  + \frac{y}{1-y} \slm_{l=0}^q  {x^\p_D}_l \right)p^+ {y^\p}_{q+1}^- }    \right] 
% \nn \\
%
% \ata 
\frac{1}{2p^+ q^- (1-y)}
\eea

\nt
In the above equation, the expression $\h( y_n^-  >  y_{n-1}^- > \cdots)$ is 
meant to indicate a product of  $n$ $\h$-functions: \tie, 

\bea
 \h( y_n^-  >  y_{n-1}^- > \cdots) =  \prod_{i=0}^{n-1} \h( y_{i+1}^-  - y_i^-  ) .
\eea
\nt
These $\h$-functions organize the longitudinal locations 
of the scatterings, starting from the hard incoming virtual photon vertex and moving towards the cut line 
(the same is true for the primed locations in the complex conjugate amplitude).  

The numerator of the momentum dependent part of Eq.~\eqref{O_00_4}, which is denoted as $N(T)$ may be 
decomposed, as in the previous section, into four parts [\tie, $ N(T) = \sum_{i=1}^4 N(T)_i $] 
depending on the components 
of the photon polarization tensor chosen. These are denoted in the same 
order as in Eq.~\eqref{w_mu_nu_twist=2}: $N(T)_1$ denotes the term arising from the projection 
$G^{\perp \perp}$. Using this combination of the components $\A,\B$, the leading part of the 
numerator may be simplified as (in the remaining equations in this section, we will ignore the vector symbol on the transverse momenta $\vl_\perp,\vp_\perp^{\,i},\vec{p^\p}_\perp^k$; they should however, always be understood to be two dimensional vectors), 

\bea
N(T)_1 \!\!\!\!&=& \!\!\!\!\tr \!\!\left[ \hf \left( \prod_{i=0}^{p-1} \g^- \g^+ q^-  \right) \g^- \left\{ \g^+ q^ - + \g^{\perp} \sum_{j=0}^{p} p^j_\perp \right\} \right. 
% \nn \\
%
% \ata 
{\g^\perp}_\A  \left\{  \g^+ ( q^ - - l^-)  + \g^{\perp} \left( \sum_{j=0}^{p} p^j_\perp - l_\perp \right)  \right\}  \nn \\
 \ata  \left( \prod_{i=p+1}^{n-1} \g^- \g^+ (q^- - l^- ) \right) \g^- \g^+ l_q^- (- g_\perp^{\A\B}) 
% \nn \\
%
% \ata
 \left( \prod_{k=m-1}^{q+1} \g^- \g^+ (q^- - l^- ) \right) \nn \\
\ata \left. \g^- \left\{  \g^+ ( q^ - - l^-)  + \g^{\perp} \left( \sum_{l=0}^{q} {p^\p}^l_\perp - l_\perp \right)  \right\}  
% \nn \\
%
% \ata 
 {\g^\perp}_\B \left\{ \g^+ q^ - + \g^{\perp} \sum_{l=0}^{q} {p^\p}_\perp \right\} 
% \nn \\
%
% \ata \left.
 \left( \prod_{k=q-1}^{0} \g^- \g^+ q^-  \right) \right].
\eea

\end{widetext}

In the above equation, the transverse components of the propagators are only retained for those propagators in the immediate 
vicinity of the photon insertion points, identified by the $\g^\perp_\A , \g^\perp_\B$ in the above equation. Only the transverse 
components in these locations give leading contributions. The transverse components in the remaining propagators yield 
suppressed contributions compared to the longitudinal components of the momenta, $q^-$ and $q^- - l^-$. 
Using the well known relations between the various $\g$ matrices, 
such as $\g^\pm \g^\pm =0 $, $\{\g^\pm , \g^\mp\} = 2 {\bf 1}$ (where ${\bf 1}$ is the 
unit matrix in spinor space) 
and the anti-commutation rule $\{\g^\pm,\g^\perp\} = 0$, the entire numerator term in the equation above may be readily simplified as, 

\bea 
N(T)_1 &=& (2q^-)^{n+m+1} (1-y)^{n+m-p-q-1}   \nn \\ 
\ata 2 \left[  \left\{ 1 + (1-y)^2 \right\} 
\left( \slm_{j=0}^p p_\perp^j \right) \x \left( \slm_{l=0}^q {p^\p}^l_\perp \right)  \right. \nn \\
\apa \left.  l_\perp^2 
- l_\perp \x \left( \slm_{j=0}^p p^j_\perp + \slm_{l=0}^q {p^\p}^l_\perp \right) \right].
\eea

\nt
Using a similar procedure as above, the terms originating from the projection $G^{+ \perp}$ and $G^{\perp +}$ are written 
in combination as (we refrain from presenting the full derivation of these terms as it is rather straightforward),

\bea
\mbx && N(T)_2 + N(T)_3  = (2q^-)^{n+m+1} \frac{(1-y)^{n+m-p-q} }{y} \nn \\
\ata 2 \left[ 2 l_\perp^2 - l_\perp \x \left( \slm_{i=0}^p p^i_\perp + \slm_{l=0}^q {p^\p}^l_\perp   \right) 
(2-y)\right]
% &-& \left.  l_\perp \x \left( \slm_{i=0}^{p} p^i_\perp  +  \slm_{l=0}^q {p^\p}^l_\perp \right)(1-y) \right] .
%
\eea

\nt
Note that these factors appear solely due to the use of the light-cone gauge. In a covariant gauge such as Feynman gauge 
there are no photon polarization tensor components which connect the $+$ and $\perp$  components.  
The last contribution originates from the projection $G^{+ +}$,  
\begin{widetext}

\bea
N(T)_4 &=& (2q^-)^{m+n+1} (1-y)^{m+n-p-q+1} 2 \frac{2 l_\perp^2}{y^2}.
\eea
\nt
The sum of these four terms has the rather simple and physical form, 
\bea
N(T) &=& (2q^-)^{n+m+1} \frac{(1-y)^{n+m-p-q -1 }}{y} 2 P_\g (y) \nn \\ 
&\times& \left[ | l_\perp |^2  - y \left\{  l_\perp \x \left( \sum_{i=0}^p p_\perp^i 
+ \sum_{k=0}^{q} {p^\p}^k_\perp  \right) \right\}  
 + y^2 \left\{ \left( \slm_{j=0}^p p_\perp^j \right) \x \left( \slm_{l=0}^q {p^\p}^l_\perp \right) \right\} \right], 
\eea
where, $P_\g(y)$ represents the photon splitting function. 

The full contribution to the term $\Op_{p,q}^{n,m}$ of Eq.~\eqref{O_00_4}, post  
integration over all longitudinal fractions, may now be reconstituted by combining 
the various parts as stated in Eq.~\eqref{general_form}:
\bea 
\Op_{p,q}^{m,n}\!\!\!  &=& \!\! e^2\! \int \!\!
\frac{d^2 l_\perp d l^-}{(2\pi)^3 2l^-}  
\frac{d^2{l_q}_\perp }{(2\pi)^2}  \prod_{i=0}^n d y_i^- d^2 y_\perp^i 
\prod_{j=1}^{m} d {y^\p}_j^- d^2 {y^\p}^j_\perp 
%\nn \\
%
% & & \int 
\prod_{i=0}^n \frac{ d^2 p^i_\perp}{(2\pi)^2} 
\prod_{j=0}^{m - 1} \frac{d^2 {p^\p}^j_\perp} { (2\pi)^2} 
(2\pi)^2 \kd^2 ( \vec{l_q}_\perp   +  \vec{l}_\perp  - \vec{K}_\perp ) \nn \\
% \nn \\
%
% 
\ata \frac{2 P_\g(y)  e^{-ix_B p^+ y_0^-} }{y  (2p^+q^-)^2 (1-y)^2 } 
 \left[ | l_\perp |^2  - y \left\{  l_\perp \x \left( \sum_{i=0}^p p_\perp^i 
+ \sum_{k=0}^{q} {p^\p}^k_\perp  \right) \right\}  
 + y^2 \left\{ \left( \slm_{j=0}^p p_\perp^j \right) \x \left( \slm_{l=0}^q {p^\p}^l_\perp \right) \right\} \right] \nn \\
 \ata \prod_{i=0}^p  e^{-i {x_D}_i p^+ y_i^-}   e^{i p^i_\perp \x  y^i_\perp  } 
\prod_{i=p+1}^{n} e^{-i  \left( \frac{{x_D}_i - \kd {x_L}_i  }{1-y}   \right)p^+  y_i^-  } 
 e^{i p^i_\perp \x  y^i_\perp  }  \prod_{i=n}^1 \h ( y_i^-   - y_{i-1}^- ) \nn \\
\ata \prod_{k=0}^{q} e^{i  {x^\p_D}_k  p^+ {y^\p}_k^- } 
e^{   -i  {p^\p}^k_\perp \x  {y^\p}^k_\perp  } 
\prod_{k=q+1}^{m} e^{ i  \left( \frac{{x^\p_D}_k - \kd {x^\p_L}_k  }{1-y}   \right)p^+  {y^\p}_k^-    } 
e^{   -i  {p^\p}^k_\perp \x  {y^\p}^k_\perp  } \prod_{k = m}^1 \h ( {y^\p}_k^-  -  { y^\p }_{ k - 1 }^- ) \nn \\
\ata \frac{1}{ {x_L}_p +  \frac{y}{1-y} \sum_{j=0}^p {x_D}_j}  
 \left[  e^{  -i \left( {x_L}_p   
+ \frac{y}{1-y} \slm_{i=0}^{p} {x_D}_i \right) p^+ y_p^-} 
% \right. \nn \\
%
% &-& \left. 
- e^{  -i \left( {x_L}_p  + \frac{y}{1-y} \slm_{i=0}^p  {x_D}_i \right)p^+ y_{p+1}^- }    \right] \nn \\
\ata \frac{1}{ {x_L^\p}_q +  \frac{y}{1-y} \sum_{l=0}^q {x^\p_D}_l}  
 \left[  e^{  i \left( {x^\p_L}_q  + \frac{y}{1-y} \slm_{l=0}^{q} {x^\p_D}_l  \right)   p^+ {y^\p}_q^-  } 
-  e^{  i \left( {x^\p_L}_q  + \frac{y}{1-y} \slm_{l=0}^q  {x^\p_D}_l \right)p^+ {y^\p}_{q+1}^- }    \right]  \nn \\
\ata \left< A; p \left| 
g^{m+n}\tr \left[ \prod_{i=1}^{n} t^{a_i}  A_{a_i}^+ (y_i^-,y^i_\perp)
%
% \ata \left. 
\prod_{j=n^\p}^{1} t^{a_j} A_{a_j}^+ ( {y^\p}_j^-, {y^\p}^j_\perp ) \right]  \right| A;p \right>.
\eea

In order to simplify the discussion of 
the remaining sections, the soft photon radiation approximation, \tie, $y \ra 0$ will be introduced. 
Only the leading behavior in $y$ will be retained. This leads to a considerable simplification 
of the various expressions and also represents a well studied limit. Two of the energy loss formalisms, 
those of Refs.~\cite{ASW} and~\cite{GLV} are cast in this limit. 
In the limit of the soft radiation approximation, all factors of $1-y$ in the denominators 
may be replaced with $1+y$ in the numerators and the leading and next-to-leading contributions in $y$ retained. 
As a result of this approximation, only the contributions from $N(T)_4$ and $N(T)_2 + N(T)_3$ need to 
be retained as they contains the leading and sub-leading negative power of $y$ and are thus dominant in the 
limit $y \ll 1$. In this limit, 

\bea
{x_L}_i  \simeq  \frac{|l_\perp|^2(1+y) - y\slm_{j=0}^{i} 2l_\perp\cdot p_\perp^i}{ 2p^+ q^- y }  
% \nn \\
\gg y {x_D}_i = y \frac{ | p^i_\perp |^2  + \slm_{j=0}^i 2 p^i_\perp \x p^j_\perp }{2 p^+ q^ - } . 
\eea

\nt
As a result, retaining solely the leading and next-to-leading behavior in $y$ we obtain $\Op_{p,q}^{m,n}$, 
in the soft radiation limit as, 
% \begin{widetext}
\bea 
\Op_{p,q}^{m,n}\!\!\!  &=& \!\! e^2\! \int \!\!
\frac{d^2 l_\perp d l^-}{(2\pi)^3 2l^-}  
\frac{d^2{l_q}_\perp }{(2\pi)^2}  \prod_{i=0}^n d y_i^- d^2 y_\perp^i 
\prod_{j=1}^{m} d {y^\p}_j^- d^2 {y^\p}^j_\perp 
%\nn \\
%
% & & \int 
\prod_{i=0}^n \frac{ d^2 p^i_\perp}{(2\pi)^2} 
\prod_{j=0}^{m - 1} \frac{d^2 {p^\p}^j_\perp} { (2\pi)^2} 
(2\pi)^2 \kd^2 ( \vec{l_q}_\perp   +  \vec{l}_\perp  - \vec{K}_\perp ) 
% \nn \\
%
% \ata 
\frac{2 P_\g(y)  e^{-ix_B p^+ y_0^-} }{y  (2p^+q^-)^2  (1 -y)^2}   \nn \\
 \ata \left[ | \vl_\perp |^2  - y   \vl_\perp \x \left( \sum_{i=0}^p \vp_\perp^i 
+ \sum_{k=0}^{q} \vec{p^\p}^k_\perp  \right)  \right]
 \prod_{i=0}^p  e^{-i {x_D}_i p^+ y_i^-}   e^{i p^i_\perp \x  y^i_\perp  } 
% \nn \\
%
%  \ata 
\prod_{i=p+1}^n e^{-i ( {x_D}_i - \kd {x_L}_i)(1+y) p^+ y_i^- } 
e^{i p^i_\perp \x  y^i_\perp  } \nn \\ 
\ata \prod_{k=0}^{q} e^{i  {x^\p_D}_k  p^+ {y^\p}_k^- } 
e^{   -i  {p^\p}^k_\perp \x  {y^\p}^k_\perp  } 
%\nn \\
%
% \ata 
\prod_{k=q+1}^{m} e^{i  (  {x^\p_D}_k  - \kd {x^\p_L}_k  ) (1+y) p^+ {y^\p}_k^- } 
e^{ -i {p^\p}^k_\perp \x  {y^\p}^k_\perp  } 
% \nn \\
%
% \ata 
\prod_{i=n}^1 \h ( y_i^-   - y_{i-1}^- ) 
\prod_{k = m}^1 \h ( {y^\p}_k^-  -  { y^\p }_{ k - 1 }^- ) \nn \\
\ata \frac{1}{{x_L}_p} \left[  e^{-i {x_L}_p p^+ y_p^- }    -   e^{ - i {x_L}_p  p^+ y_{p+1 }^-}\right]  
%\nn \\
%
% \ata 
\frac{1}{{x_L^\p}_q} \left[  e^{i {x_L^\p}_q p^+ { y_q^\p }^- } 
 -   e^{ i { x_L^\p }_p  p^+ { y_{q+1}^\p }^-}\right]  \nn \\
\ata \left< A; p \left| 
g^{m+n}\tr \left[ \prod_{i=1}^{n} t^{a_i}  A_{a_i}^+ (y_i^-,y^i_\perp)
%
% \ata \left. 
\prod_{j=n^\p}^{1} t^{a_j} A_{a_j}^+ ( {y^\p}_j^-, {y^\p}^j_\perp ) \right]  \right| A;p \right>.
\label{O_m_n_simple}
\eea
\end{widetext}
\nt
The expression derived above is completely general, in the sense that no assumption 
regarding the nature of the nuclear state has been made. In the next section, a re-summation 
over the different locations of where the photon is radiated will be carried out. Following which, 
a factorization of the above hadronic tensor into a part that is solely dependent on hard momenta and 
a part dependent on soft momenta will be carried out and simplifying assumptions regarding 
the nuclear state made. This will then be followed by the re-summation over number of 
scatterings. 

%%%%%%%%%%%%%%%%%%%%%%%%%%%%%%%%%%%%%%%%%%%%%%%%%%%%%%%%%%%%%%%%%%%%%%%
%%%%%%%%%%%%%%%%%%%%%%%%%%%%%%%%%%%%%%%%%%%%%%%%%%%%%%%%%%%%%%%%%%%%%%%
%%%%%%%%%%%%%%%%%%%%%%%%%%%%%%%%%%%%%%%%%%%%%%%%%%%%%%%%%%%%%%%%%%%%%%%
%%%%%%%%%%%%%%%%%%%%%%%%%%%%%%%%%%%%%%%%%%%%%%%%%%%%%%%%%%%%%%%%%%%%%%%

\section{Sum over photon production points, factorization and gradient expansion}

%%%%%%%%%%%%%%%%%%%%%%%%%%%%%%%%%%%%%%%%%%%%%%%%%%%%%%%%%%%%%%%%%%%%%%%
%%%%%%%%%%%%%%%%%%%%%%%%%%%%%%%%%%%%%%%%%%%%%%%%%%%%%%%%%%%%%%%%%%%%%%%
%%%%%%%%%%%%%%%%%%%%%%%%%%%%%%%%%%%%%%%%%%%%%%%%%%%%%%%%%%%%%%%%%%%%%%%
%%%%%%%%%%%%%%%%%%%%%%%%%%%%%%%%%%%%%%%%%%%%%%%%%%%%%%%%%%%%%%%%%%%%%%%

In the preceding section, a general expression for the single photon bremsstrahlung 
from a multiply scattering hard quark parton was derived. The expression in Eq.~\eqref{O_m_n_simple} represents the 
case where the quark scatters $n$ times in the amplitude and $m$ times in the complex conjugate. 
The photon is produced after the $p^{\rm th}$ scattering in the amplitude and after the $q^{\rm th}$
scattering in the complex conjugate. To obtain the full differential cross section to produce a photon with 
momentum fraction $y$ and transverse momentum $l_\perp$, sums have to be carried out over the 
various quantities $p,q$ and $n,m$. A few comments are in order: in analogy with the case of 
transverse momentum broadening of a hard quark propagating in a nuclear medium~\cite{Majumder:2007hx}, 
cases where $n \neq m$ represent unitarity corrections to terms where the produced quark scatters $\min[n,m]$
times. The cases where $p=q$ represent squares of the amplitude where the photon is produced in the $p^{\rm th}$
location. The cases where $p\neq q$ represent interference terms which lead to the well known 
Landau-Pomeranchuck-Migdal (LPM) suppression of the photon production rate.

In this section, the sum over the various locations of the radiated photon in the amplitude and the 
complex conjugate amplitude (\tie, $p,q$ ) will be carried out. Invoking the small $y$ approximation, 
the $l_\perp$ dependent momentum fractions may be expanded as, 

\bea
{x_L}_p &= & x_L(1+y) - \sum_{i=0}^{p}\kd {x_L}_i \nn \\
\aqa  \frac{l_\perp^2 (1+y)}{2p^+q^- y}  
- \frac{\sum_{i=0}^{p}  2 l_\perp \x p^i_\perp }{2 p^+ q^- }
\eea

\nt
To simplify notation, we set $\D x_L^i = \sum_{j=0}^i \kd {x_L}_j$.  
As most of the factors in $\Op_{p,q}^{m,n}$  [Eq.~\eqref{O_m_n_simple} ] are independent 
of  $p,q$, this contribution may be written compactly as,
\begin{widetext}
\bea
\Op_{pq}^{mn} &=& \int \frac{d^2 l_\perp d l^-}{(2\pi)^3 2l^-}  
\frac{ d^2 {l_q}_\perp}{(2\pi)^2} \md (y^-,y_\perp)  \md ( {y^\p}^- , y^\p_\perp )  
\md p_\perp \md p^\p_\perp  
%\nn \\
%
% \ata 
 M(\fy,\fy^\p)  (2\pi)^2  \kd^2 ( \vec{l_q}_\perp   +  \vec{l}_\perp  - \vec{K}_\perp ) 
 \frac{2 e^{-ix_B p^+ y_0^-} }{y  (2p^+q^-)^2 (1-y)^2} \nn \\
\ata P_\g(y)   \prod_{i=0}^n e^{-ix_D^i p^+ y_i^- + i p_\perp^i \x y_\perp^i}
\prod_{j=0}^m e^{-i{x^\p_D}^i p^+ {y^\p}_i^- + i {p^\p_\perp}^i \x {y^\p_\perp}^i} 
% \nn \\
%
% \ata 
 f_p g_q , \label{general_form_2}
\eea

\nt
where, the terms that depend on the location where the photon is produced are denoted separately. The 
product of these factors is given, in the small y limit as, 

\bea
f_p g_q\!\!\!\aqa \!\!\!\frac{1 - y + \frac{\D x_L^p}{x_L} }{x_L} \left[   
e^{-i [(x_L(1+y)  -  \D x_L^p  ] p^+  y_p^-   +\, i \!\!\!\!\slm_{i=p+1}^n\!\!\!\!  \kd {x_L}_i p^+ y_i^-    } 
% \right.  \nn \\
%
% &-& \left.  
\!\!\!\!\!\!-  e^{-i  [ x_L(1+y)  -  \D x_L^p  ]  p^+ y_{p+1}^-   + \,i \!\!\!\!\slm_{i=p+1}^n \!\!\! \kd {x_L}_i p^+ y_i^-    }  \right]  
\left( \vl_\perp - y\sum_{i=0}^p  \vec{p^i}_\perp \right) \nn  \\ 
\ata \!\!\!\!\frac{1 - y + \frac{\D {x^\p}_L^q  }{  x_L } }{x_L} \left[   
e^{+i  [ x_L(1+y)  -  \D {x^\p}_L^q  ] p^+ {y^\p}_q^-   -\,  i \!\!\!\!\slm_{j=q+1}^m  \!\!\! \kd {x^\p_L}_j p^+ {y^\p}_j^-    } 
% \right.  \nn \\
%
% &-& \left.  
\!\!\!\!-  e^{+i  [ x_L(1+y)  -  \D {x^\p}_L^q  ] p^+ {y^\p}_{q+1}^-   
-\, i \!\!\!\!\slm_{j=q+1}^m \!\!\!  \kd {x^\p_L}_j p^+ {y^\p}_j^-    }  \right] 
\left( \vl_\perp - y\sum_{k=0}^q  \vec{p^\p}^k_\perp \right). \nn
\label{f_p_g_q}
\eea
 \nt 
In both the denominators and in the exponent, the leading contribution is from the term $x_L \sim 1/y$. The 
remaining terms $\D x_L^p , \kd {x_L}_i $ and $x_D^i $ are all independent of, hence subleading in, $y$. Therefore,  
the factor of $y$ in products such as $\kd {x_L}_i (1+y)$ may be dropped.

The sums to be carried out are those over $0<p<n-1$ and $0<q<m-1$ restricted to the terms $f_p g_q$. The 
reader will note that these two sums are independent of each other and as a result may be carried 
out rather easily. The sum over $p$ is 

\bea
\!\!\!\!\sum_{p=0}^{n-1} \frac{1 - y+ \frac{\D x_L^p }{ x_L} }{x_L} \left[   
e^{-i  [ x_L(1+y)  -  \D x_L^p  ]p^+ y_p^-   + \,\,i\!\!\! \slm_{i=p+1}^n  \kd {x_L}_i p^+ y_i^-    } 
% \right.  \nn \\
%
% &-& \left.  
\!\!\!\!\!\!-  e^{-i  [x_L(1+y)  -  \D x_L^p  ] p^+ y_{p+1}^-   +  \,\,i\!\!\! \slm_{i=p+1}^n  \kd {x_L}_i p^+ y_i^-    }  \right] 
\left( \vl_\perp - y\sum_{i=0}^p  \vec{p^i}_\perp \right).
\eea

\nt
Note that the $2^{nd}$  phase factor in the $p^{\rm th}$ term is identical to the $1^{st}$ phase factor in the 
$(p+1)^{\rm th}$ term. As a result, many terms cancel in the sum if the small correction 
$\D x_L^p $ in the coefficient of the phases is neglected. The phases which never cancel are the 
$1^{st}$ phase of the term $p=0$ and the $2^{nd}$ phase of the term $p=n-1$.  The result of the 
sum over $p$, up to next-to-leading terms in $y$, is 

\bea 
\sum_{p=0}^{n-1} f_p  \aqa 
 \frac{1}{x_L}  \left[ (1-y)\vl_\perp \left\{ e^{-i x_L(1+y) p^+ y_0^- } e^{i \sum_{i=0}^n  \kd {x_L}_i p^+ y_i^-  }  
% \right.   \\
%
% &-&  
-  e^{-i x_L(1+y) p^+ y_n^- }   e^{i \sum_{i=0}^n\kd {x_L}_i p^+ y_n^-  }  \right\} \right. \nn  \\
&+&  \sum_{p=0}^{n-1}\left(  \frac{\vl_\perp \kd {x_L}_p }{x_L} - y \vec{p}^{\,p}_\perp \right)
e^{-i  (x_L(1+y)  -  \D x_L^p  )p^+ y_p^-   + i \sum_{i=p+1}^n  \kd {x_L}_i p^+ y_i^- }   \nn \\
\apa \left. \sum_{i=0}^{n-1} \left(  \frac{\vl_\perp \kd {x_L}_i }{x_L} - y \vec{p}^{\,i}_\perp \right) 
e^{-i  (x_L(1+y)  -  \D x_L^{n}  )p^+ y_n^-  } \right].
\eea

\nt
In the limit that the photon is extremely soft, one may retain solely the leading contribution in 
$y$, \tie, only the first two terms in the brackets above. This corresponds to the ``deep-LPM'' limit:
The regime where the radiated photon cannot resolve the various scatterings and treats the entire 
nucleus (or the entire process of $n$ scatterings) coherently as one single scattering event. The 
next-to-leading term in $y$ which is the third term in the equation above represents the 
first correction to this limit where the photon just begins to resolve the different scatterings of the 
hard quark in the medium. The terms obtained from the sum $\sum_{q=0}^{m-1}$ in the 
complex conjugate amplitude yield a near 
identical expression with $n$ replaced by $m$ and the phase factors complex conjugated.

We may now add the terms $f_n$ and $g_m$ \tie, the cases where the photon is produced after all the 
$n$ or $m$ scatterings.  Such contributions necessarily have an on-shell, cut quark line on one side of the 
photon vertex. As a result, these terms do not contain the difference of two phase factors as in 
Eq.~\eqref{photon_vertex_props} and only have the first factor which occurs when the line after the 
photon emission is taken on shell (in this case taken by cutting the line). For the sum over $p$ this 
yields,

\bea 
\sum_{p=0}^{n-1} f_p  \aqa 
 \frac{1}{x_L}  \left[ (1-y)\vl_\perp e^{-i x_L(1+y) p^+ y_0^- } e^{i \sum_{i=0}^n  \kd {x_L}_i p^+ y_i^-  } \right. \\
%
% &-&  
&+&  \left.  \sum_{p=0}^{n-1}\left(  \frac{\vl_\perp \kd {x_L}_p }{x_L} - y \vec{p}^{\,p}_\perp \right)
e^{-i  (x_L(1+y)  -  \D x_L^p  )p^+ y_p^-   + i \sum_{i=p+1}^n  \kd {x_L}_i p^+ y_i^- }    
 + \left(  \frac{\vl_\perp \kd {x_L}_n }{x_L} - y \vec{p}^{\,n}_\perp \right)
e^{-i  (x_L(1+y)  -  \D x_L^n  )p^+ y_n^-  } \right] .  \nn 
\eea

\end{widetext}

We will also ignore 
the next-to-leading terms in $y$ that occur in  the phases as these will not play a major role in what follows. 
With these simplifications, we obtain 

\bea 
\sum_{p,q=0}^{n,m} f_p g_q \aqa
\frac{e^{-i x_L p^+ y_0^- }}{x_L^2} l^2_\perp \nn \\
\ata \left[  1 - 2y + y \sum_{p=1}^{n} \frac{ p^p_\perp \x l_\perp }{ l_\perp^2} 
e^{-i  x_L  p^+ (y_p^- - y_0^-) }   \right.  \nn \\
\apa  \left.  y \sum_{q=1}^{m} \frac{ {p^\p}^q_\perp \x l_\perp }{ l_\perp^2 } 
e^{ i  x_L  p^+ {y^\p}_q^-  }   \right].  \label{fpgp_final}
\eea

Up to this point, the collinear approximation has been used to simplify the 
expressions for the hadronic tensor, without the introduction of factorization. 
The separation of the hadronic tensor into a hard short distance piece and a soft 
long distance contribution may now be accomplished. All factors in Eq.~\eqref{O_m_n_simple} 
which contain the hard scales $p^+,q^-$ constitute the hard part. All factors that
depend solely on the soft $\perp$ momenta and distances along with the matrix element 
constitute the long distance element. 
All phase 
factors which contain a factor $ x_D^i p^+ $ (as well as its primed counterparts) or $ x_L p^+$ 
as part of their arguments belong in the  hard part. 
The purely transverse phase factors such as $\exp [ i \vec{p}^i_\perp \x \vec{y}^i_\perp] $  (which do not 
contain any factor of $l_\perp$) belong in the 
soft part along with the matrix elements. Thus, we may decompose the general contribution to the hadronic tensor as, 

\bea 
\sum_{p,q}^{n,m} \Op_{p,q}^{n,m} = \int \md y \md p_\perp H(p^+,q^-,p_\perp,l_\perp,y)  S(p_\perp, y) . \label{W_mu_nu_simple}
\eea
\nt
In the above equation, $y$ and $p_\perp$ are representative of the entire set of distances and transverse 
momentum that appear in Eq.~\eqref{O_m_n_simple}. 

The soft part which contains the matrix elements of the gluon vector potentials in the nuclear state may be  
simplified first. The simplifications arise from approximations made regarding the nature of the nuclear state $| A; p\rc$.
In this manuscript, the nucleus is approximated as a weakly interacting homogeneous gas of nucleons. 
Such an approximation is only sensible at very high energy, where, due to 
time dilation, the nucleons appear to travel in straight lines almost independent of each other 
over the interval of the interaction of the hard probe. In a sense, all forms of correlators  
between nucleons (spin, momentum, etc.) are assumed to be rather suppressed. 
As a result, the expectation of 
the $n+n^\p+2$ operators in the nuclear state may be decomposed as 

\bea
&& \lc A;p | \psibar(y^-,y_\perp)\g^+ \psi(0) \prod_{i=1}^{n+n^\p} A^{+}_{a_i}(y_i)| A; p\rc \nn \\
&=& A C^A_{p_1}  \lc p_1 | \psibar(y^-,y_\perp)\g^+ \psi(0) \prod_{i=1}^{n+n^\p} A^{+}_{a_i}(y_i)  | p_1 \rc
\nn \\
&+& C^A_{p_1,p_2}  \lc p_1| \psibar(y^-,y_\perp)\g^+ \psi(0) | p_1 \rc \nn \\
\ata \lc p_2 | \prod_{i=1}^{n+n^\p} A^{+}_{a_i}(y_i)  | p_2 \rc + \ldots , \label{operators_nuclear_state}
\eea
\nt 
where, the factor $C^A_{p_1}$ represents the probability to find a nucleon in the vicinity of the location $\vec{y}$, which is a 
number of order unity (it is the probability to find one of  $A$ nucleons distributed in a volume of size $c A$ within a nucleon 
size sphere centered at $\vec{y}$). The remaining coefficients $C^A_{p_1,\ldots}$ represent the weak 
position correlations between different nucleons. The overall factor of $A$ arises from the determination of the origin (the location $0$ in 
the equation above) in the nucleus, 
which may be situated on any of the $A$ nucleons. 
Solely for the current discussion, we reintroduce the quark operators $\psibar(y^-,y_\perp)$ and $\psi(0)$ in the 
above equation. 

It is clear from the 
above decomposition that the largest contribution arises from the term where the expectation of 
each partonic operator is evaluated in separate nucleon states as the $\vec{y}_i$ integrations 
may be carried out over the nuclear volume. 
As a nucleon is a color singlet, any combination of quark or gluon field strength insertions in 
a nucleon state  must itself  be restricted to a  color singlet combination. As a result, the expectation of 
single partonic operators in nucleon states is vanishing. The first (and hence largest) non-zero contribution 
emanates from the terms where the quark operators in the singlet color combination are evaluated in a 
nucleon state and the $n+n^\p$ gluons are divided into pairs of singlet combinations, with each singlet 
pair evaluated in a separate nucleon state. This requires that $n+n^\p$ is even and may lead to 
a maximum overall factor of 

\bea
C^A_{p_1,p_2,\ldots}  \sim  A^{[(n+n^\p+2)/2] } ,
\eea
\nt
in the large $A$ limit.  It should be pointed out that large contributions may also arise, in principle, 
when $n+n^\p$ is odd. In this case, the two quarks and a gluon are considered in the singlet 
combination with the remaining 
gluons evaluated in singlet pairs in the remaining nucleons. Here we institute the experimental 
observation that $(n)$-parton observables are much smaller than $(n-1)$-parton observables. 
This is only true, once again, outside the saturation regime. 
In this effort, the focus remains exclusively outside this region, as a result we ignore all terms 
with more than two quarks or two gluons per nucleon. 

Further simplifications arise in the evaluation of  gluon pairs in a singlet combination in the nucleon 
states by carrying out the $y_\perp$ integrations. The basic object under consideration is  (ignoring the 
longitudinal positions and color indices on the vector potentials)

\bea
\mbx & & \int  d^2 y^i_\perp  d^2 {y^\p}^j_\perp \lc p | A^{+}  (\vec{y}^i_\perp )  A^{+}  ( \vec{y^\p}^j_\perp ) | p \rc \nn \\
&\times & e^{-i x_D^i p^+ y_i^-}  e^{ i p^i_\perp \x y^i_\perp}   
e^{i {x^\p}_D^j p^+ {y^\p}_j^-}  e^{ - i {p^\p}^j_\perp \x {y^\p}^j_\perp}  \nn \\
&=& (2\pi)^2 \kd^2( {\vp}^{\,i}_\perp - {\vec{p^\p}} ^j_\perp )  \int d^2 y_\perp  e^{-i x_D^i p^+ ( y_i^-  -  {y^\p}_j^- )} \nn \\
% \frac{-g^{\A \B}_{\perp}}{2} \nn \\ 
%
&\times& 
 e^{ i p_\perp \x y_\perp} \lc p | A^{+} (\vec{y}_\perp/2 )  A^{+}  ( - \vec{y}_\perp/2 ) | p \rc , \label{two_gluon_cor}
 \eea 
where, $y_\perp$ is the transverse gap between the two gluon insertions and $p_\perp = (p^i_\perp + p^j_\perp)/2$. 
The physics of the 
above equation is essentially the transverse translation symmetry of the two gluon correlator 
in a very large nucleus. One will note that the two dimensional delta function over the transverse 
momenta has removed an integration over the transverse area of the nucleus thus reducing the 
overall $A$ enhancement that may be obtained. This is then used to equate the transverse 
momenta emanating from the two gluon insertions in the amplitude and complex conjugate amplitude. 
This also simplifies the longitudinal phase factors which now depends solely on the difference of the 
longitudinal positions of the two gluon insertions. In order to simplify further discussion, we consider the 
specific case where there are $n$ gluon insertions in the amplitude and $m=n+2\D n$ insertions in the 
complex conjugate (where $\D n$ is a positive integer).  
Under the assumptions of restricting each nucleon 
state to be acted upon by only two gluon operators, we find that $n+\D n$ nucleons are involved.
The $n$ path ordered gluon insertions in the amplitude are matched up by a series of path ordered 
insertions in the complex conjugate amplitude on the same set of nucleons. The remaining $\D n$ nucleons 
with 2 gluon insertions each may be distributed between the $n$ nucleons. The equating of transverse momenta 
emanating from and being injected into these $\D n$ nucleons insists that they impart vanishing net transverse 
momentum to the propagating quark. 
As such, the transverse momenta emanating from such nucleons does not contribute to the overall transverse 
broadening of the quark and so does not have any appearance in the overall transverse momentum delta function.
Thus the overall transverse momentum $\kd$-function only contains $n$ different transverse momentum 
variables from $p_\perp^1$ to $p_\perp^n$.

One now invokes the collinear 
approximation in expanding the hard part as a Taylor expansion in transverse 
momenta around the origin $p_\perp^i \ra 0$. In the case of a cut with $m=n+2\D n$ there are 
$2n$ gluon insertions which produce transverse momentum broadening and as a result, as many 
derivatives, which involve $n$ different transverse momenta.
All terms of the form 
\[
 \prod_{i=1}^p \frac{1}{2} \frac{\prt^2  }{ \prt {p_\perp^i}^\A \prt {p_\perp^i}^\B} 
\left.  H \right|_{p_\perp = 0}{p_\perp^i}^\A {p_\perp^i}^\B  ,
\]
where, $p<n$, yield gauge corrections for the contributions with $2p$ 
gluon insertions and as many derivatives~\cite{lqs}. The genuine $2n$ twist correction at this 
order has $p=n$.  Expanding to this order, one obtains the 
generic term,

\bea
& & \prod_{i=1}^n \frac{1}{2} \frac{\prt^2  }{ \prt {p_\perp^i}^\A \prt {p_\perp^i}^\B} 
\left. H(p^+, q^-, p_\perp^i, )\right|_{p_\perp^i = 0 } \nn \\
\ata {p_\perp^i}^\A {p_\perp^i}^\B   A^+_{a_i} (y_i^-,y_\perp^i/2)
 A^+_{a^\p_j} (  {y^\p}_j^-   ,   -y_\perp^i/2   ). \label{2n_der}
\eea
\nt
Where, we have assumed the result of Eq.~\eqref{two_gluon_cor} and reintroduced the 
color indices and longitudinal locations.

Using integration by parts over the transverse distance $y_\perp^i$ one may convert 
the product ${p_\perp^i}^\A A^+_{a_i} (y_i^-,y_\perp^i/2) \ra \frac{1}{2} \prt_\perp^\A A^+_{a_i} (y_i^-,y_\perp^i/2)$ .
In the extreme collinear limit, in the presence of a hard scale such that $g$ is small, one may make the approximation,

\bea 
\prt_\perp A^+_a \simeq {F_a}^+_\perp, \label{glue_field}
\eea
where, ${F_a}^+_\perp$ represents the gluon field strength.  Carrying this out consistently on the 
two gluon operator in the nucleon state of Eq.~\eqref{two_gluon_cor} and ignoring derivatives of 
the field strength [$\prt^\A F^{+ \B} \sim g(m^2) g^{\A \B} j^+ \ra 0$] we obtain, 
\bea
\mbx \!\!\!\!\!\!\!\!\!\!\!\! && 
\int d^2 y_\perp {p_\perp^i}^\A {p_\perp^i}^\B  e^{ i p_\perp \x y_\perp} \lc p | A^{+} (\vec{y}_\perp/2 )  A^{+}  ( - \vec{y}_\perp/2 ) | p \rc  \nn \\
\mbx \!\!\!\!\!\!\!\!\!\!\!\! &=& 
\int d^2 y_\perp  e^{ i p_\perp \x y_\perp} \frac{1}{2} \lc p | F^{+ \A} (\vec{y}_\perp/2 )  F^{+ \B}  ( - \vec{y}_\perp/2 ) | p \rc \nn \\
\mbx \!\!\!\!\!\!\!\!\!\!\!\!&=& 
\int d^2 y_\perp  e^{ i p_\perp \x y_\perp} \frac{-g_\perp^{\A\B}}{4} \lc p | F^{+ \rho} (\vec{y}_\perp/2 )  F^{+}_{\rho , }  ( - \vec{y}_\perp/2 ) | p \rc.
 \eea 

\nt
In the last line of the above equation we have averaged over the spins in the two field strength expection in the nucleon state with the constraint that 
the operator being evaluated in the nucleon be a spin singlet. The nucleon states are always assumed to be spin singlets or in spin averaged states.

With the derivative 
expansion (at vanishing transverse momenta $p^i_\perp \ra 0$) imposed on the hard part $H$, it no longer has any functional 
dependence on the transverse momenta. The integrations over the transverse 
momenta may now be included completely into the soft part. 
As in the case of transverse broadening~\cite{Majumder:2007hx}, the action of the transverse 
momentum derivatives on the phase factors will not be considered. All such derivatives 
necessarily extract a factor of spatial separation $y_j$ (with $0<j<n,m$), as 
\bea
\frac{\prt}{\prt p^j_\perp} e^{- i {x_D}_j p^+ y_j^- } = 
-i y_j^- \frac{\prt {x_D}_j }{\prt p^j_\perp} p^+ e^{- i {x_D}_j p^+ y_j^- }.
\eea
These result in spatial moments of the two gluon matrix elements such as 
$\lc p | F^{+  \rho } (y_j^-, y^j_\perp )  y_j^- F^{\,\,+}_{ \rho, } (0 ) | p \rc$, which will be ignored in this effort.

The action of the $2n$ derivatives is thus restricted to the overall transverse momentum conserving delta 
function and the transverse momentum  factors of Eq.~\eqref{fpgp_final}. The remnant transverse momentum 
dependence lies in the soft part which contains the matrix elements of the gluon field strengths and 
\mbox{$\kd$-functions} which equate pairs of transverse momentum. 
We point out that, while we are considering cases where $n = m -2 \D n$ the final results for $n = m + 2\D n$ are 
identical. 
The entire structure of the term
with $n+m$ gluon insertions and $2n$ transverse derivatives may be expressed as 
(we also approximate $\frac{1}{(1-y)^2} \sim 1 + 2y$ ), 

\begin{widetext}
\bea 
\Op^{n,m}  \!\!\!\!&=&\!\!\!\!  \sum_{p,q}^{n,m}   \Op_{p,q}^{n,m}   
 = \int \frac{d^2 l_\perp d y}{(2\pi)^3  y }  
\frac{ d^2 {l_q}_\perp}{(2\pi)^2} 
\md y^-  \md  {y^\p}^-  
% \nn \\
%
% && 
 \md p_\perp \md p^\p_\perp 
\frac{P_\g(y) }{ y (2 p^+ q^-)^2} e^{ -i (x_B + x_L) p^+ y_0^- } \label{O_nm} \\
\ata \!\!\!  \left\{ \prod_{j=1}^{n}  \frac{ \prt^2 }{ \prt {p_j}^{\A_j}_\perp  \prt {p_j}^{\B_j}_\perp } \right\} \left[  
\left\{ \prod_{i=1}^{n}  \h( y_i^-  -   y_{i-1}^- ) e^{-i {x_D}_i p^+ y_i^- }  \right\}
%\\
%
%\ata 
\left\{ \prod_{k=1}^{m} \h( {y^\p}_k^-  -   { y^\p }_{k-1}^- )  e^{ i {x_D}_k p^+ {y^\p_k}^- } \right\} \right.  \nn \\
\ata \!\!\!\! \left. 
 \frac{l_\perp^2}{x_L^2}\left\{1  + y \sum_{i=1}^n \frac{  {p_i}_\perp  \x l_\perp }{ l_\perp^2 }  e^{-ix_L p^+ (y_i^-  - y_0^- ) } 
 + y \sum_{k=1}^{m} \frac{  {p^\p_k}_\perp \x  l_\perp   }{ l_\perp^2 }  e^{  i x_L p^+ {y^\p}_k^-  } \right\}
\kd^2( l_\perp + {l_q}_\perp - \slm_{q=0}^n {p_q}_\perp )\right] \nn \\
\ata C^A_{p_1,\ldots,p_{n-\D n}  }  \mbx^{n-1}P_{\D n}  
\left\{ \prod_{i}^{n  } (2 \pi)^2 \kd^2 ( p^i_\perp - {p^\p}^i_\perp ) \int d^2 y^i_\perp 
\lc p_i | F^{+ \mu_i} F^+_{\mu_i} | p_i \rc e^{i  \bar{p}_\perp^i \x  y^i_\perp} \frac{(- g_{\perp}^{\A_i \B_i} ) }{8}  \right\} \nn \\
\ata \left\{ \prod_{l=0,2,4, \ldots}^{2\D n}   \!\!\!\!\!\!\!\!{\bf \p}  (2 \pi)^2 \kd^2 ( {p^\p}^l_\perp + {p^\p}^{l+1}_\perp ) \int d^2 {y^\p}^l_\perp 
\lc p_l | A^+({y^\p_l}^-, y^l_\perp/2 ) A^+({y^\p_{l+1}}^-, - y^l_\perp/2) | p_l \rc e^{i \kd {p^\p}_\perp^l \x  y^l_\perp} \right\} . \nn
\eea
% \end{widetext}
In the above equation, $^{n-1}P_{\D n}$ represents the number of permutations of $\D n$ pairs of consecutive 
gluon insertions in $n-1$ locations. The \emph{barred} transverse momenta $\bar{p}_\perp = ( p_\perp + p^\p_\perp )/2 $, 
whereas $\kd {p^\p_l}_\perp = {p^\p_l}_\perp - {p^\p_{l+1}}_\perp$ represent the consecutive gluon pair insertions with two 
gluons per nucleon.
The transverse momentum $\delta$-functions in the soft part that 
correspond to the second type of insertions require that the momentum brought in by one such gluon is taken out 
immediately by the other. As a result, such insertions play no role in the double differential transverse 
momentum distribution being calculated and may be ignored. They however provide unitarity corrections 
to the total cross section. Alternatively, one may state that such insertions induce two exactly opposing 
currents in close proximity and thus produce a vanishing net transverse broadening and completely 
destructive radiative contributions.  In the remainder we will consider only symmetric contributions, 
where $n=m$ or $\D n = 0$.

In the symmetric case, one may carry out ${p^\p}^k_\perp$ integrations using 
the transverse momentum $\delta$-functions in the 
soft part. As a result, there is a pair-wise equality between the transverse momentum brought in 
by the gluon insertions on the left and right hand side of the cuts. Using integration by parts one may 
also replace $\prt/\prt {p^\p}^k_\perp \ra - \prt/\prt p^i_\perp $.
Setting $n=m$ in Eq.~\eqref{O_nm}, the action of  the derivatives $\prt/\prt p_{i \perp}$ (where $1<i<n$)  on the 
transverse momentum dependence in the hard part may be simplified as, 

\bea
&& \prod_{i=1}^n (-g_\perp^{ \A_i  \B_i  }  ) \frac{ \prt }{  \prt  { p^i_\perp }^{ \A_i }  } 
\frac{ \prt }{ \prt  { p^i_\perp }^{ \B_i }  } \left[  
\kd^2 ( l_\perp + {l_q}_\perp - \sum_{i=0}^{n}  p^i_\perp )  
% \right. \nn \\
%
% \ata 
\frac{e^{-i x_L p^+ y_0^- }}{x_L^2} \right. \nn \\ 
\ata \left.  \left\{  1  + 
%\\
 y \sum_{p=1}^{n}\frac{ p^p_\perp \x l_\perp }{ l_\perp^2} 
 \left( e^{-i  x_L  p^+ (y_p^- - y_0^-) }  
% \right.  \nn \\
%
% \apa 
+  
% \frac{ p^p_\perp \x l_\perp }{ l_\perp^2 } 
%
e^{ i  x_L  p^+ {y^\p}_p^-  } \right) \right\} \right] ,
%
% \apa y^2  \left( \sum_{p=1}^{n} \frac{ p^p_\perp \x l_\perp }{ l_\perp^2} 
% %
% e^{-i  x_L  p^+ (y_p^- - y_0^-) }  \right) \nn \\ 
% %\\
% \ata \left. \left. \left(  \sum_{q=1}^{n} \frac{ p^q_\perp \x l_\perp }{ l_\perp^2 } 
% %
% e^{ i  x_L  p^+ {y^\p}_q^-  }    \right)  \right\}  \right],
\eea

\nt where the transverse derivatives may act entirely on the transverse momentum $\delta$-function or 
one or two derivatives may act on the terms within the curly brackets above and the remaining may 
act on the $\delta$-function. The $p^i_\perp$ derivatives acting on the transverse momentum 
$\kd$-function may be replaced as  

\bea
\frac{\prt}{\prt p^j_\perp} \kd^2  (l_\perp + l_{q _\perp} - \sum_{i=1}^n p_{i_\perp}) 
%\nn \\ 
%
=  - \frac{\prt}{\prt l_{q, \perp}} \kd^2 (l_\perp + l_{q _\perp} - \sum_{i=1}^n p_{i_\perp}). 
% \nn
%
\eea
\nt
Using the above equation, the action of the $2n$ derivatives on the combination of 
the two dimensional transverse momentum dependent $\delta$-function and the phase 
factor $\sum_{p,q} f_p g_q$ followed by an imposition of the limit of very small transverse 
momenta $p^i_\perp \ra 0$ may be simply expressed as, 

%%%%%%%%%%%%%%%%%%%%%%%%%%%%%%%%%%%%%%%%%%%%
%%%%%%%%%%%%%%%%%%%%%%%%%%%%%%%%%%%%%%%%%%%%
%%%%%%%%%%%%%%%%%%%%%%%%%%%%%%%%%%%%%%%%%%%%
%%%%%%%%%%%%%%%%%%%%%%%%%%%%%%%%%%%%%%%%%%%%
%%%%%%%%%%%%%%%%%%%%%%%%%%%%%%%%%%%%%%%%%%%%
%%%%%%%%%%%%%%%%%%%%%%%%%%%%%%%%%%%%%%%%%%%%

\bea
&& \left. \prod_i  \frac{\prt^2}{\prt {p^i_\perp}^2} 
\kd^2 (\vec{l}_\perp + \vec{l_q}_\perp - \vec{K}_\perp)  \sum_{p,q} f_p g_q \right|_{p_\perp = 0} \label{n_equal_nprime} \\
&=&  \frac{e^{-i x_L p^+ y_0^- }}{x_L^2} l_\perp^2 \left[   \left( \nabla_{l_{q_\perp}}^2 \right)^n 
% \right. \nn \\
%
% &-& 
\!\!\!- y \sum_{p=1}^{n} \left( e^{-i  x_L  p^+ (y_p^- - y_0^-) }  +  e^{i  x_L  p^+ {y^\p}_p^- } \right)
\frac{ l_\perp \x \nabla_{l_{q_\perp}} }{ l_\perp^2} 
\left( \nabla_{l_{q_\perp}}^2 \right)^{n-1} \right] 
\kd^2 (\vec{l}_\perp + \vec{l_q}_\perp - \vec{K}_\perp). \nn
\eea
\end{widetext}

%%%%%%%%%%%%%%%%%%%%%%%%%%%%%%%%%%%%%%%%%%%%
%%%%%%%%%%%%%%%%%%%%%%%%%%%%%%%%%%%%%%%%%%%%
%%%%%%%%%%%%%%%%%%%%%%%%%%%%%%%%%%%%%%%%%%%%
%%%%%%%%%%%%%%%%%%%%%%%%%%%%%%%%%%%%%%%%%%%%

The action of the derivatives on the transverse $\kd$-function or on the 
factors of $p_\perp^i$ that appear in $f_p g_q$ are 
the only non-vanishing contributions from Eq.~\eqref{O_m_n_simple}.
The longitudinal integrals, due to color 
confinement,  yield the requirement that the longitudinal locations of the two gluons which act on the same 
nucleon state be in close 
proximity. The factor of $\kd y_p^-$ represents the small gap between the longitudinal positions between the 
gluon insertions in a single nucleon.  
One now tries to identify the most length enhanced term by isolating the maximum number of 
unconstrained $dy^-$ integrals.  Note that, due to the assumption of short distance color correlation
 (ignoring color and spin indices), 

\bea
\int d  y^- d {y^\p}^- \lc p | F  (y^- )  F ({y^\p}^- ) | p \rc  \simeq \int d y^-  \lc F F \rc y^-_c,  \label{correlation_length}
\eea
where, $\lc FF \rc$ is the gluon expectation at the mean location $y^- \pm y^-_c$ and $y^-_{c}$ represents the color 
correlation length  in the medium. In a nucleus, this is equivalent to the confining distance, whereas 
in a quark gluon plasma it would be related to the Debye length. 

Each such integral yields a factor of $L^- \sim A^{1/3}$ from the unconstrained $y^-$ integration.  
Equating the pairs of transverse momenta 
that appear in each two-gluon correlation, as 
well as using the relation between the longitudinal momenta from the $\h$-functions in Eq.~\eqref{O_m_n_simple}, 
require that the largest length enhancement arises from the terms 
where the gluon correlations are built up in a mirror symmetric fashion, \tie, where the gluon insertion at $y^i$ is 
contracted with that at ${y^\p}^i$.  
One may now let the transverse momenta $p^i_\perp$ (for all $i$) in the hard part 
tend to zero and integrate over the remaining $p^i_\perp$ integrals in the soft part, setting the corresponding transverse 
distances between the gluon insertions in a nucleon to zero \tie, $y^i_\perp \ra 0$.

One may average colors of the quark and gluon field operators, 

\bea 
\lc p | F^a F^b | p \rc = \frac{\kd^{ab}}{ (N_c^2 - 1)} \lc p | F^a F^a | p \rc \label{glue_color_conf}\\ 
\lc p | \psibar_i\g^+ \psi_j | p \rc = \frac{\kd_{ij}}{ N_c } \lc p |  \psibar \g^+ \psi| p \rc. \label{quark_color_conf}
\eea 
\nt
This reduces the overall trace over color factors to 
\bea
&& \frac{1}{N_c (N_c^2 - 1)^n} \tr \left[ \prod_{i=1}^n t^{a_i} \prod_{j=n}^1 t^{a_j} \right] \nn \\
&=& \frac{C_F^n}{(N_c^2 - 1)^n}
= \frac{1}{(2N_c)^n}. 
\eea
\nt

The remaining $n$ longitudinal position integrals for the gluon insertions may be simplified as 

\bea
\int \prod_{i=1}^{n} dy_i^-  \h(y_i^- - y_{i-1}^-) = \frac{1}{n!} \int \prod_{i=1}^n dy_i^- .
\eea
\nt
Invoking the above simplifications, the leading length enhanced contribution at order $2n$ to the 
term $\Op^{nn}$ is obtained as,

\bea 
\Op^{n,n} \!\!\!\!&=&\!\!\! \frac{\A_{em }}{2\pi}\int \frac{d l^2_\perp d y}{ l_\perp^2}  d^2{l_q}_\perp 
P_\g(y) e^{-i( x_B + x_L)p^+ y_0^-} \label{O_2n_final} \\
\ata \!\!\!\!C^A_{p_0,\ldots, p^n} \left[  \frac{ \left( \bar{D}L^- \nabla_{l_{q_\perp}}^2 \right)^n }{n!}  
 - y \left\{ \bar{E}^+ (x_L)  \right. \right.\nn \\ 
\apa \!\!\!\!\bar{E}^- (x_L) \left. \left. \right\}  
%
% \ata \!\!\!\!\left. 
 \frac{ l_\perp \x \nabla_{l_{q_\perp}} }{ l_\perp^2} 
\frac{ \left( \bar{D}L^- \nabla_{l_{q_\perp}}^2 \right)^{n-1} }{(n-1)!} \right] \kd^2 (\vec{l}_\perp + \vec{l_q}_\perp ) .  \nn
\eea

\nt
In the above equation, the function
$\bar{D}$, represents the following expectation value in a nucleon state, 

\bea
 \bar{D} \aqa \frac{\pi^2 \A_s}{2N_c} 
 \int \frac{dy^-}{2\pi} \lc p | {F^a}^{+ \A}(y^-) {F^a}_{\A,}^{\,\,\, +}(0) | p \rc ,  
\eea
\nt
While no limits have been specified in the integration over longitudinal separation $y^-$, 
it should be understood that due to the assumption of a short distance color correlation length 
in the medium [as in Eq.~\eqref{correlation_length}], the integral in the above equation only 
receives strength form the region with $| y^- | \leq y_c $. 
The \emph{nuclear} functions, $ \bar{E}^\pm$ represent the following expectation value in the nuclear state, 
\bea
\bar{E}^\pm (x_L) \aqa \frac{\pi^2 \A_s}{2N_c}  \int \frac{dy^- d y_p^-}{2\pi}e^{ \pm i  x_L  p^+ (y_p^-  -  y_0^-) }  \nn \\ 
\ata  c_{OF} \lc p | {F^a}^{+ \A}(y_p^- + y^-) {F^a}_{\A,}^{\,\,\, +}(y_p^-) | p \rc .  \label{E1}
\eea
\nt
As in the case of the correlator $\bar{D}$, the integration over the longitudinal separation $y^-$ is limited by the 
color correlation length in the medium or the confining distance in a nucleus. The integration over the mean location 
of the two gluon field strength insertions \tie, $y_p^-$ has its upper bound constrained only by the size of the medium 
$L$ (in the case of a nucleus $L \sim A^{1/3}$). The lower bound of $y_p^-$, in the case of $E^+$, is given by $y^-$: 
the location of the $\psibar$ operator insertion in Eq.~\eqref{operators_nuclear_state}. In the case of $E^-$, the lower 
bound is $y_p^- = 0$, given by the location of the $\psi$ operator in Eq.~\eqref{operators_nuclear_state}. It should 
be pointed out the two lower bounds are separated by at most one unit of the color correlation length (the size of a nucleon in 
the case of DIS on a large nucleus) and as such, this small distinction will be ignored in what follows.  

While the two nucleon states on either side in the above equation are denoted as carrying the same momentum $p$, this 
may not strictly be the case. Throughout this calculation, the hard scattering piece, which includes the scattering of the 
initial virtual photon with the incoming quark (including the nucleon which contained the quark) was factored out [see Eq.~\eqref{hard_fact}]. 
In the case of the first term in Eq.~\eqref{O_2n_final}, which represents the case that the photon is radiated immediately 
after the hard scattering, such a factorization along with the deconvolution of the nucleus into nucleon states is well justified 
as there is no longitudinal momentum which is picked up in the multiple scattering which leads to the transverse 
broadening. In the case that the photon is radiated at a later scattering, the nucleon state which is struck by the 
hard virtual photon and the nucleon state which radiates the final outgoing photon may exchange longitudinal 
momentum differently between them in the amplitude and in the complex conjugate. In such a case, these two 
nucleon states represent the convolution of two generalized parton distribution (GPD) functions~\cite{Ji:1996ek}, 
where the excess momentum from one such distribution is balanced by the other. The momentum correlation between these two 
states, no doubt, depends on the distance $y_p^-$ between them. The overall constant  $c_{OF}$ in Eq.~\eqref{E1} 
accounts for both the average momentum correlation in a large nucleus and corrections resulting from the approximation 
of replacing a generalized parton distribution by a regular diagonal parton distribution. A phenomenological discussion 
on $c_{OF}$ and its value in large nuclei may be found in Refs.~\cite{Osborne:2002st}.
In the next section, a re-summation of all such contributions of arbitrary order $n$ will be carried out.

%%%%%%%%%%%%%%%%%%%%%%%%%%%%%%%%%%%%%%%%%%%%%%%%%%%%%%%%%%
%%%%%%%%%%%%%%%%%%%%%%%%%%%%%%%%%%%%%%%%%%%%%%%%%%%%%%%%%%
%%%%%%%%%%%%%%%%%%%%%%%%%%%%%%%%%%%%%%%%%%%%%%%%%%%%%%%%%%
%%%%%%%%%%%%%%%%%%%%%%%%%%%%%%%%%%%%%%%%%%%%%%%%%%%%%%%%%%
%%%%%%%%%%%%%%%%%%%%%%%%%%%%%%%%%%%%%%%%%%%%%%%%%%%%%%%%%%

  \section{Re-summation and soft photon production from multiple scattering}
% 
% %%%%%%%%%%%%%%%%%%%%%%%%%%%%%%%%%%%%%%%%%%%%%%%%%%%%%%%%%%
% %%%%%%%%%%%%%%%%%%%%%%%%%%%%%%%%%%%%%%%%%%%%%%%%%%%%%%%%%%
% %%%%%%%%%%%%%%%%%%%%%%%%%%%%%%%%%%%%%%%%%%%%%%%%%%%%%%%%%%
% %%%%%%%%%%%%%%%%%%%%%%%%%%%%%%%%%%%%%%%%%%%%%%%%%%%%%%%%%%
% 

In the preceding section, the length enhanced contributions of twist-$2n$ to the soft 
photon double differential rate were extracted. It turned out that the photon rate at twist-$2n$ 
(including the leading and next-to-leading contributions in $y$) depended on two different 
correlation functions $\bar{D}$ and $\bar{E}$ evaluated in the 
nuclear state, as well as on the multi-nucleon combinatorial factor $C^A_{p_0,\ldots,p_n}$.  
In the case where terms such as these are not very small compared to the leading twist 
contribution, all such terms need to be re-summed.

In order to 
sum over all $n$, this last factor has to be simplified. In a sense, re-summation requires that this coefficient 
have a formally multiplicative structure.  
In the preceding section, 
a model of the nucleus as a weakly interacting homogeneous gas of nucleons was used. The 
formal expression  of this assumption is hidden within the dimensionful parameter $C^A_{p_0,\ldots,p^n}$. 
The precise evaluation of such combinatorial coefficients is rather complicated, even for the case of 
next-to-leading twist~\cite{Osborne:2002st}.  From general dimensional arguments, in the case of 
non-interacting nucleons, this factor may be approximated as, 

\bea
C^A_{p_0,\ldots, p^n} \simeq C^A_p  \left( \frac{\rho}{2 p^+} \right)^n,
\eea
\nt
where, $\rho$ is the nucleon density inside the nucleus and $1/2p^+$ originates in the 
normalization of  the nucleon state. The remaining unknown coefficient $C^A_p$ 
is now considered to be independent of the order $n$ and is included in  the leading 
twist hadronic tensor of Eq.~\eqref{w_mu_nu_twist=2}.  This decomposition is 
also identical to that used in the case of transverse broadening. It should be pointed 
out that there may still be an extra normalization factor for each nucleon and this 
may be included with the factor of $\rho/2p^+$. We will ignore any such normalization 
factors, as they are not essential to the re-summation. It should however be pointed out that the
correlation between the nucleons off which the hard parton scatters and the original nucleon 
struck by the hard virtual photon is different from that between the original nucleon and that 
where the outgoing photon is emitted. As mentioned above, these constitute off-forward 
distributions where there is momentum shared between the two nucleon states. 

The decomposition of the combinatorial factor allows for the definition of the new quantities, 
\bea
D = \frac{\rho}{2p^+} \bar{D} = \frac{\pi^2 \A_s}{2 N_c} \rho \int \frac{dy^-}{2\pi 2 p^+} \lc p | {F^a}^{+ \A} {F^a}_{\A,}^{\,\,\, +}| p \rc, \label{D}
\eea
and similarly,
\bea
E^{\pm} (x_L) = \frac{\rho}{2p^+} \bar{E}^{\pm} .
\eea
Note that $D$ is exactly the diffusion tensor for the transverse broadening experienced by a hard parton 
traversing the nucleus without radiation. This is directly proportional to the well known jet transport parameter 
$\hat{q}$~\cite{Majumder:2007hx,Majumder:2006wi}:

\bea
\hat{q} = \frac{2 \lc l_\perp^2 \rc_{L^-}}{L^-} = 8 D.  
\eea
\nt
The new quantities $E^{\pm}$ may be related to $D$ as,
\bea
E^{\pm} (x_L) = \int d y_p^- c_{OF} e^{ \pm i x_L p^+ ( y_p^-   -   y_0^- ) } D(y_p^-).
\eea 
It should be pointed out that $D$ is a local quantity evaluated at particular location $y_p$ and
could very well change with time in a non-static medium, whereas the $E$'s are intergrated over the 
space time path of the jet as it traverses the medium. 

In the following, we will only present the re-summation of the quantity $\Op^{n,n}$, \tie,
\bea
\Op = \sum_{n=0}^{\infty} \Op^{n,n}.
\eea
The re-summed differential hadronic tensor and the differential cross section may be easily obtained 
by incorporation of the above quantity into Eqs.~(\ref{LO_cross},\ref{w_mu_nu_twist=2}). 
The final state operator matrix element for a parton undergoing $1<n<\infty$ scatterings may be re-expressed as,
\bea 
\sum_{n=0}^\infty \Op^{n,n} &=& \frac{\A_{em }}{2\pi} \int 
\frac{d l^2_\perp d y}{ l_\perp^2}  d^2{l_q}_\perp 
P_\g(y) e^{-i( x_B + x_L)p^+ y_0^-}  \nn \\
\ata \sum_{n=0}^\infty  \left[  \left\{  \frac{ \left( D L^- \nabla_{l_{q_\perp}}^2 \right)^n }{n!}  \right\}   
\kd^2 (\vec{l}_\perp + \vec{l_q}_\perp )  \right. \nn \\
&-& c_p \left\{ y \left[  E^+ (x_L)  + E^- (x_L) \right] \frac{ l_\perp \x \nabla_{l_{q_\perp}} }{ l_\perp^2} \right\} \nn \\
\ata \left.  \frac{ \left( D L^- \nabla_{l_{q_\perp}}^2 \right)^{n-1} }{(n-1)!} 
\kd^2 (\vec{l}_\perp + \vec{l_q}_\perp ) \right] .   \label{O_2n_sum}
\eea
In the above equation, it is understood that the second term in the square bracket does not receive 
contributions from the case of $n=0$. The reader may immediately verify that for the case of 
$n=0$, which corresponds to the case of no scattering, the first term by itself reproduces the result of 
Eq.~\eqref{O_00_3} for the case of $y \ra 0$. The new coefficient $c_p$ accounts for the weak 
correlation between the nucleon struck by the hard virtual photon and the nucleon from which 
the outgoing photon is radiated. In principle, $c_p$ may depend on the shared 
momentum fraction $x_L$. 

Both the terms in the above equation may be re-summed with the observation that the sum, 
\[
\phi(l_{q_\perp},L^-) = \sum_{n=0}^\infty 
\frac{ \left( D L^- \nabla_{l_{q_\perp}}^2 \right)^n }{n!} \kd^2 (\vec{l}_\perp + \vec{l_q}_\perp ),
\]
obeys the diffusion equation, 
\bea
\frac{\prt}{\prt L^-} \phi(l_{q_\perp},L^-) = D \nabla_{l_{q_\perp}}^2  \phi(l_{q_\perp},L^-). \label{diffusion_eqn}
\eea
with the initial condition discerned from Eq.~\eqref{d_W_0} as, 
\bea
\phi(L^-=0, \vec{l}_{q_\perp}) = \kd^2(\vec{l}_{q_\perp} + \vec{l}_\perp ). \label{initial_cond}
\eea 
As demonstrated in the case of transverse broadening, contributions such as those from Eq.~\eqref{O_nm} 
lead to a unitarization of the cross sections and, as a result, normalize the solutions of the diffusion equation above. 
The general normalized solution to Eq~\eqref{diffusion_eqn} is given as~\cite{sneddon},
\bea 
\phi(L^-,\vec{l}_\perp) = \frac{1}{4 \pi D L^-} 
\exp \left\{-  \frac{\left| \vec{l}_\perp + \vec{l_q}_\perp \right|^2}{4 D L^-} \right\}. \label{solution}
\eea
The reader will note, that $\phi(L^-,l_\perp)$ reconverts back to the two dimensional delta 
function in the limit of $L^- \ra 0$ and the solution is unitary in the sense that 
\[
 \frac{\prt}{\prt L^-} \int d^2  l_\perp \phi(l_\perp,L^-) \simeq 0. 
 \]

Substitution of the above solution back into Eq.~\eqref{O_2n_sum}, yields the very simple expression 
for the final state re-summed operator matrix element $\Op$, \tie,
\bea
\Op &=& \frac{\A_{em }}{2\pi}\int 
\frac{dl^2_\perp d y}{ l_\perp^2} d^2{l_q}_\perp 
 P_\g(y) e^{-i( x_B + x_L)p^+ y_0^-}  \nn \\
\ata \left[ 1  + y c_p \frac{\left\{ E^+ (x_L) + E^- (x_L) \right\} }{2 D L^-} 
\frac{l_\perp^2 + \vec{l}_\perp \x \vec{l_q}_\perp}{l_\perp^2} \right] \nn \\
\ata \phi(L^-,l_{q_\perp}).  \label{rate}
\eea

\nt
Substitution of the above final state operator matrix element into Eq.~\eqref{w_mu_nu_twist=2}, 
yields the full differential hadronic tensor and as a result, the differential cross section for 
photon production at all twist. 

\bea
\frac{d {W^A}^{\mu \nu}}{dy dl_\perp^2 d^2 {l_q}_\perp} &=& C_p^A 2 \pi 
 \sum_q Q_q^4 
(- g_\perp^{\mu \nu}) 
%
% \ata  
\frac{\A_{em}}{2\pi}  \frac{ P_{q\ra q \g }(y) }{l_\perp^2} \nn \\
\ata \int \frac{dy_0^-}{2\pi}  e^{-i(x_B + x_L)p^+ y_0^-} F_q (y_0^-)  \nn \\
\ata \left[\frac{\mbx}{\mbx} 1  
+ yc_p \frac{\left\{ E^+ (x_L) + E^- (x_L) \right\} }{2 D L^-} \right. \nn \\
\ata \left. \frac{l_\perp^2 + \vec{l}_\perp \x \vec{l_q}_\perp}{l_\perp^2} \right] 
\phi(L^-,l_{q_\perp}).   \label{mult_scat_diff_rate}
\eea
\nt
The above equation represents the main result of this article. The unintegrated 
expectation value of the two quark operator $F(y_0)$ is defined in Eq.~\eqref{F(y_0)}.
The differential 
spectrum of photons radiated from a hard parton undergoing multiple scattering 
in the medium has been expressed as a factorized product of the photon 
splitting function, the two dimensional transverse momentum distribution of the 
propagating parton and a multiplicative factor 
[factor in square brackets in Eq.~\eqref{mult_scat_diff_rate}] 
which encodes the phases picked 
up by the particles as they traverse the medium. It is this factor which has to be 
evaluated for each path taken by the hard quark in the medium, weighted by the 
expectation of the two gluon field strength operator product along the path. 
The unknown coefficients, $C^A_p$ and 
$c_p$ have to be determined by a phenomenological analysis. The decomposition 
of the above equation into quark and gluon structure functions in the nucleus and 
its associated phenomenological implications will be discussed in an upcoming 
publication.

%%%%%%%%%%%%%%%%%%%%%%%%%%%%%%%%%%%%%%%%%%%%%%%%%%%%%%%%%%
%%%%%%%%%%%%%%%%%%%%%%%%%%%%%%%%%%%%%%%%%%%%%%%%%%%%%%%%%%
%%%%%%%%%%%%%%%%%%%%%%%%%%%%%%%%%%%%%%%%%%%%%%%%%%%%%%%%%%
%%%%%%%%%%%%%%%%%%%%%%%%%%%%%%%%%%%%%%%%%%%%%%%%%%%%%%%%%%
%%%%%%%%%%%%%%%%%%%%%%%%%%%%%%%%%%%%%%%%%%%%%%%%%%%%%%%%%%

\section{Conclusions and Discussions}

%%%%%%%%%%%%%%%%%%%%%%%%%%%%%%%%%%%%%%%%%%%%%%%%%%%%%%%%%%
%%%%%%%%%%%%%%%%%%%%%%%%%%%%%%%%%%%%%%%%%%%%%%%%%%%%%%%%%%
%%%%%%%%%%%%%%%%%%%%%%%%%%%%%%%%%%%%%%%%%%%%%%%%%%%%%%%%%%
%%%%%%%%%%%%%%%%%%%%%%%%%%%%%%%%%%%%%%%%%%%%%%%%%%%%%%%%%%
%%%%%%%%%%%%%%%%%%%%%%%%%%%%%%%%%%%%%%%%%%%%%%%%%%%%%%%%%%

The study of the modification of hard jets in  dense matter is now approaching a rather sophisticated stage 
of its development. The wide variety of experimental measurements of the modification of jets and jet like correlations 
require the emergence of a single formalism capable of describing the modification of hard jets over a wide range 
of energies and a variety of different media. In this manuscript, the higher-twist mechanism of jet modification 
is extended in an effort to make it applicable to both the \emph{thin} medium and \emph{thick} medium limits. 
The limits of  thick and thin refer to the number of scatterings that a hard parton will encounter on its path 
through the medium. While previous efforts~\cite{HT,Guo:2006kz} have focused on the modification due to a few scatterings, 
the current manuscript furthers the development of the formalism required to describe the modification due to an infinite 
number of scatterings. 

The focus of the current manuscript has been to compute the spectrum of single photon 
bremsstrahlung from a propagating parton undergoing multiple scattering in an extended medium. 
The entire process is cast in a Deep-Inelastic Scattering framework, where a virtual photon (with virtuality $Q^2$)  
strikes a hard parton in a nucleon, which is itself  
situated in a large nucleus. The parton, on its way through the nucleus, scatters multiple times and 
radiates a real photon. Energy loss due to real gluon emission was ignored.  Each scattering 
of the hard parton is suppressed by powers of $Q^2$, but is enhanced by the length of the nucleus ( $\propto A^{1/3}$) 
over which the scattering may occur. For large $A$, an infinite  class of such contributions need to be re-summed.

The photon may be radiated from the vicinity of any of the scattering locations. The sum over such contributions, 
leads to the destructive interference for very forward radiation known as the LPM effect. In the extreme case of  
an on-shell parton entering such a medium and radiating a very soft photon with momentum fraction $y<<1$, 
it is shown that the destructive interference is 
complete. This is referred to as the deep-LPM limit. In the current effort, the first two corrections in terms of 
initial virtuality and expansion in powers of $y$,  to this limit, are 
computed. The first correction is due to the fact the the initial parton that enters the medium is not on-shell but may 
indeed be very virtual. As a result, it may radiate the photon even without the need for further re-scattering. The 
second contribution originates from not assuming the very soft limit $y \ra 0$ for the photon, instead keeping the 
first leading corrections in $y$. This is the reason our results differ from the case of no re-scattering by a factor 
proportional to $y$. 

As in the case of transverse momentum broadening, it is demonstrated that the infinite series of power corrections 
may indeed be re-summed. The result has the simple and intuitive structure of a product  of the photon 
splitting function, the two dimensional transverse momentum distribution of the 
propagating parton and a multiplicative factor which encodes the phases picked 
up by the particles as they traverse the medium. 
As in the case of multiple scattering without radiation, a Gaussian
profile in transverse momentum centered around the origin was obtained for the propagating parton~\cite{Majumder:2007hx}. 
However, the Gaussian obtained in this case was found centered around the momentum $-\vl_\perp$, needed
to balance the transverse momentum of the photon.

 The results of such a 
computation are crucial to the understanding of both the near side and away side correlations between a trigger 
hadron and an associated photon or vice-verse. The phenomenological comparisons with experiment and other theoretical 
approaches, 
based on the expressions calculated above, will appear in a future publication. The derived results also bear 
considerable relevance as a secondary source of single hard photon production.

An obvious next step is the evaluation of gluon radiation in the
same process. Ultimately this might lead the way to a unified 
description of multiple scattering in the medium and radiative 
energy loss.
While the photon does not scatter in the 
medium, its formation incorporates many similar physics issues such as the destructive interference 
of the LPM effect. Such effects will reappear in the future calculation of gluon radiation and will have to be dealt 
with in the case of energy loss of hard partons. However, the various scales that will be encountered in that 
problem as well as the approximation scheme used will be near identical as that in the current 
manuscript.

%%%%%%%%%%%%%%%%%%%%%%%%%%%%%%%%%%%%%%%%%%%%%%%%%%%%%%%%%%
%%%%%%%%%%%%%%%%%%%%%%%%%%%%%%%%%%%%%%%%%%%%%%%%%%%%%%%%%%
%%%%%%%%%%%%%%%%%%%%%%%%%%%%%%%%%%%%%%%%%%%%%%%%%%%%%%%%%%
%%%%%%%%%%%%%%%%%%%%%%%%%%%%%%%%%%%%%%%%%%%%%%%%%%%%%%%%%%
%%%%%%%%%%%%%%%%%%%%%%%%%%%%%%%%%%%%%%%%%%%%%%%%%%%%%%%%%%

\section{Acknowledgments}

%%%%%%%%%%%%%%%%%%%%%%%%%%%%%%%%%%%%%%%%%%%%%%%%%%%%%%%%%%
%%%%%%%%%%%%%%%%%%%%%%%%%%%%%%%%%%%%%%%%%%%%%%%%%%%%%%%%%%
%%%%%%%%%%%%%%%%%%%%%%%%%%%%%%%%%%%%%%%%%%%%%%%%%%%%%%%%%%
%%%%%%%%%%%%%%%%%%%%%%%%%%%%%%%%%%%%%%%%%%%%%%%%%%%%%%%%%%
%%%%%%%%%%%%%%%%%%%%%%%%%%%%%%%%%%%%%%%%%%%%%%%%%%%%%%%%%%

This work was supported in part by the U.S. Department of Energy
under grants DE-FG02-05ER41367, DE-AC02-98CH10886, RIKEN/BNL, and the
Texas A\&M College of Science. A.~M. thanks the Cyclotron Institute and Department 
of Physics, Texas A\&M University, where this effort was begun, for their kind hospitality. 
A.~M. and R.~J.~F. also thank Charles Gale for discussions and the Department of Physics, McGill 
University for kind hospitality.


\begin{thebibliography}{99}
\bibitem{f2} 
A more correct notation would be to refer 
to this state as $| A- 1\rc$, ignoring the $1$ is appropriate for large $A$. These issues are dealt with more rigor in Sect.~IV.
\end{thebibliography}
\end{document}